\documentclass[twocolumn, trackchanges]{aastex61}
\shortauthors{Plotkin et al.}

\begin{document}

\title{Radio Variability from a Quiescent Stellar Mass Black Hole Jet}
\shorttitle{Quiescent Black Hole Radio Variability}

\author[0000-0002-7092-0326]{R.~M.~Plotkin}
\affiliation{International Centre for Radio Astronomy Research - Curtin University, GPO Box U1987, Perth, WA 6845, Australia}

\author[0000-0003-3124-2814]{J.~C.~A.~Miller-Jones}
\affiliation{International Centre for Radio Astronomy Research - Curtin University, GPO Box U1987, Perth, WA 6845, Australia}

\author[0000-0002-8400-3705]{L. Chomiuk}
\affiliation{Department of Physics and Astronomy, Michigan State University, East Lansing, MI 48824, USA}

\author[0000-0002-1468-9668]{J. Strader}
\affiliation{Department of Physics and Astronomy, Michigan State University, East Lansing, MI 48824, USA}

\author{S. Bruzewski}
\affiliation{Department of Physics and Astronomy, Michigan State University, East Lansing, MI 48824, USA}
\affiliation{Department of Physics and Astronomy, University of New Mexico, 800 Yale Blvd NE, Albuquerque, NM 87131, USA}

\author{A. Bundas}
\affiliation{Department of Physics and Astronomy, Michigan State University, East Lansing, MI 48824, USA}
\affiliation{Jennifer Chalsty Planetarium, Liberty Science Center, Jersey City NJ, USA}

\author{K. R. Smith}
\affiliation{International Centre for Radio Astronomy Research - Curtin University, GPO Box U1987, Perth, WA 6845, Australia}

\author[0000-0001-8665-5523]{J. J. Ruan}
\affiliation{McGill Space Institute and Department of Physics, McGill University, 3600 rue University, Montreal, QC H3A 2T8, Canada}

\correspondingauthor{R. M. Plotkin}
\email{richard.plotkin@curtin.edu.au}

\newcommand{\ledd}{L_{\rm Edd}}   
\newcommand{\ergs}{{\rm erg~s}^{-1}}

\newcommand{\lr}{L_{\rm R}}   
\newcommand{\lx}{L_{\rm X}}     

\newcommand{\xrb}{BHXB}
\newcommand{\tmsp}{J1023}

\newcommand{\src}{V404 Cygni}
\newcommand{\phcal}{J2025+337}

\newcommand{\rsquo}{'}

\newcommand{\nvla}{127}   
\newcommand{\nvlaWithBad}{129}
\newcommand{\nvlaXwithBad}{86}
\newcommand{\nvlax}{84}   
\newcommand{\nvlac}{24}   
\newcommand{\nvlau}{11}   
\newcommand{\nvlak}{4}     
\newcommand{\nvlal}{4}      
\newcommand{\njvla}{5}   
\newcommand{\ndecay}{4}  

\newcommand{\nvlba}{14}
\newcommand{\nvlbax}{1}
\newcommand{\nvlbac}{13}

\newcommand{\nvlaxtot}{89}   

\newcommand{\nallobs}{150}
\newcommand{\nstatscband}{38}   
\newcommand{\nstatscbandDet}{28}  
\newcommand{\nstatscbandLim}{10}  
\newcommand{\nstatsxband}{86}    
\newcommand{\nstatsxbandDet}{61} 
\newcommand{\nstatsxbandLim}{25}  
\newcommand{\nstatsxbandDetVlaonly}{60}   
\newcommand{\nstatsxbandDof}{60}   
\newcommand{\nstatsxbandDofVlaonly}{59}  

\newcommand{\note}[1]{\authorcomment1{#1}}


\begin{abstract}
Relativistic outflows are believed to be a common feature of  black hole X-ray binaries at the lowest accretion rates, when they are in their `quiescent' spectral state.  However, we still lack a detailed understanding of how quiescent jet emission varies with time.   Here we present 24 years of archival radio observations (from the Very Large Array and the Very Long Baseline Array) of the black hole X-ray binary V404 Cygni in quiescence (totalling 150 observations from 1.4 -- 22 GHz). The observed flux densities  follow lognormal distributions with means and standard deviations of $\left(\left<\log f_\nu \right>, \sigma_{\rm \log f_\nu}\right) = \left(-0.53, 0.19\right)$ and $\left(-0.53,  0.30\right)$ at  4.9 and 8.4 GHz, respectively (where $f_\nu$ is the flux density in units of mJy).  As expected, the average radio spectrum is flat with a mean and standard deviation of  $\left(\left<\alpha_r \right>, \sigma_{\alpha_r}\right)= \left(0.02, 0.65\right)$ where $f_\nu \propto \nu^{\alpha_r}$.  We find that radio flares that increase the flux density by factors of 2 -- 4 over timescales as short as $<$10 min are commonplace, and that long-term variations (over 10--4000 day timescales) are consistent with shot noise impulses that decay to stochastic variations on timescales $\lesssim$10 days (and perhaps as short as tens of minutes to several hours).  We briefly compare the variability characteristics of \src\  to jetted active galactic nuclei, and we conclude with recommendations on how to account for variability when placing quiescent black hole X-ray binary candidates with radio luminosities comparable to \src\ ($\lr \approx 10^{28}~\ergs$) onto the radio/X-ray luminosity plane.  

\end{abstract}


\section{Introduction}
\label{sec:intro}


 Black hole X-ray binaries (\xrb s) that accrete  at low Eddington ratios ($\lesssim 0.01 \ledd$, where $\ledd$ is the Eddington luminosity) are associated with compact radio emission from partially self-absorbed synchrotron jets  \citep{blandford79, fender01, remillard06}.  These relativistic outflows provide channels for  accretion flows to shed angular momentum, and to transport energy  out to large distances \citep[e.g.,][]{meier01, fender16, romero17, douna18}.   Radio observations of relativistic jets are  therefore crucial  for understanding how matter is transported through, and away from,  accretion flows with low mass accretion rates.
 
The most weakly accreting black holes  reside in the `quiescent' spectral state, which we define here by a soft X-ray spectrum that can be characterised by a power-law  photon index\footnote{The photon index $\Gamma$ is defined by $N_E \propto E^{-\Gamma}$, where $N_E$ is the photon number  density per unit energy, $E$.}  
of $\Gamma \sim 2.1$ \citep[e.g.,][]{tomsick01, kong02, corbel06, plotkin13, reynolds14}.  Most \xrb s spend the majority of their time in quiescence, which usually corresponds to (0.5-10 keV) X-ray luminosities  $\lx \lesssim 10^{-6} - 10^{-5} \ledd$ \citep{plotkin13, plotkin17}.  In quiescence,  a larger fraction of the radiative power emitted by the accretion flow/jet system appears to be emitted in the radio waveband \citep[e.g.,][although see \citealt{yuan05} for a prediction otherwise]{fender03, corbel13, gallo18}.   This increased  dominance of  radio emission   implies that the  radio domain can be effective for discovering quiescent \xrb s \citep{maccarone05, fender13}, which would produce less biased samples of \xrb s in the Milky Way compared to the traditional method of discovering \xrb s through X-ray emission during an outburst.

 Coordinated radio and X-ray surveys are indeed starting to reveal \textit{candidate} \xrb s in Milky Way globular clusters \citep{strader12, chomiuk13, miller-jones15, shishkovsky18} and in the field \citep{tetarenko16}.  Such an approach is highly complementary to other strategies for discovering quiescent \xrb s, through, e.g., H$\alpha$ surveys \citep{casares18},  X-ray surveys \citep[e.g.,][]{agol02, jonker14},  and  optical spectroscopic searches capable of discovering  \textit{non-accreting} black hole candidates in detached binary systems \citep{giesers18, thompson18}.

  At the moment, even our most sensitive radio facilities are  only capable of probing the tip of the quiescent \xrb\ population.    We have currently detected radio emission from only four nearby quiescent \xrb s ($\lesssim$4 kpc), including one of the most luminous known quiescent systems, \src\ ($\lr \approx 10^{28}\,\ergs$ at 5 GHz; \citealt{hynes04, gallo05, rana16}), and three of the least luminous known systems  (A 0620$-$00, XTE J1118+480, and MWC 656; $\lr \approx 10^{26}\,\ergs$; \citealt{gallo06, gallo14, dzib15, ribo17, dincer18}).   Thus, we are still establishing the empirical properties of quiescent \xrb\ radio jets, which is an essential step for eventually defining reliable radio-based selection criteria. 
  
  To help inform future radio surveys,  this paper focuses on quantifying the radio variability characteristics  of quiescent jets.   Understanding that variability level is important, as (i) it would establish whether radio variability can discriminate quiescent  \xrb s from other classes of radio-emitting objects;  and (ii) it would  help determine how close in time radio observations must be coordinated with other multiwavelength data.

To open the quiescent radio time domain we focus on  \src, which represents the luminous end of the quiescent \xrb\ population.    \src\ contains a dynamically confirmed black hole ($9.0^{+0.2}_{-0.6} M_\odot$) orbiting a K3 III companion  \citep{khargharia10} in a long $6.473 \pm 0.001$ day orbit \citep{casares92}, and it was the first transient BHXB to have an unambiguous X-ray detection in quiescence ($4\times10^{33}\,\ergs$ from 0.2-2.4 keV with the \textit{ROSAT} satellite; \citealt{wagner94}).  It also has a well-established distance ($2.39\pm 0.14$ kpc) from radio parallax measurements \citep{miller-jones09}, which has also been measured in the optical (albeit with poorer precision) by  \textit{Gaia} \citep{gaia-collaboration18}.   Crucially, there already  exist $\approx$10$^2$  radio observations of \src\ in the Very Large Array (VLA) archive dating back to the 1990s.  \src\ is the \textit{only} quiescent \xrb\ with such a rich radio dataset (the next best sampled quiescent \xrb s are A0620$-$00 and MWC 656, each of which have 2-3 published radio detections; \citealt{gallo06, dzib15, ribo17, dincer18}).  However, the full VLA archive has yet to be synthesized into a single variability study.   The quiescent X-ray variability characteristics of \src\ have already been well-quantified from minute through month timescales \citep[e.g.,][]{wagner94, hynes04, bradley07, bernardini14, rana16},  making \src\ ripe for radio-to-X-ray comparisons.

 In this paper we reanalyse all  observations of \src\ in quiescence with the VLA through 2015, and we also consider 14 observations with the Very Long Baseline Array (VLBA).  We describe our data reductions in Section~\ref{sec:obs}.  In Section \ref{sec:res} we describe the flux density variability characteristics on long  (days through decades) and short (minutes to hours) timescales,  and we explore radio spectra on long and short timescales.   We discuss the radio jet properties in Section~\ref{sec:disc}, and in Section~\ref{sec:mwcoord} we provide recommendations on how to combat radio variability when coordinating multiwavelength observing campaigns on quiescent \xrb s and on \xrb\ candidates at luminosities comparable to \src.  

\section{Archival Observations and Data Analysis}
\label{sec:obs}

\src\ has undergone two major outbursts during the VLA era, the first discovered  on 1989 May 22 \citep{makino89}, and the second outburst on 2015 June 15 \citep{barthelmy15, kuulkers15, negoro15, younes15}.  After the main 2015 outburst ended in July (see next paragraph), renewed X-ray activity was observed on 2015 December 21 \citep{malyshev15}, and \src\ went through a mini-outburst that lasted $\sim$30 days  \citep[see, e.g.,][]{kimura17, munoz-darias17, tetarenko19}.

For the 2015 outburst, from \citet{plotkin17} we consider that \src\ re-entered quiescence around 2015 July 23, based on when the X-ray spectrum finished softening from $\Gamma \sim 1.6$ to $\Gamma \sim 2.0-2.1$ (see their Figure 1 and Table 3). We therefore exclude VLA observations on or before July 23 in this study, but we include VLA observations after July 23 (four total).  Since the X-ray characteristics after July 23 appear similar to those before the outburst, we infer that the physical properties of the underlying accretion flow are no different during these 2015 observations  compared to  pre-outburst (and by extension, we do not expect physical differences in the radio jet pre- and post-outburst).  We therefore include these four observations from 2015 for completeness.  However, out of caution, we generally use different symbols/colors to mark  the post-outburst observations in figures within this manuscript, and we often remove the post-outburst observations from statistical tests.  We do not include any data during or after the 2015 mini-outburst.

The 1989 outburst lasted much longer than the 2015 outburst \citep[e.g.,][]{tetarenko19}.  Similar  quality X-ray spectral coverage of the transition into quiescence is  not available from 1989, making it difficult to pinpoint when  \src\ re-entered quiescence.   \citet{han92} monitored the 1989 outburst and decay with the VLA for two years, and from their Table 1 the 4.9 and 8.4 GHz flux densities remained brighter than 1 mJy even as late as 1990 September, perhaps indicating  elevated accretion  (for reference, typical flux densities between outbursts were 0.2--0.4 mJy, e.g., \citealt{gallo05, hynes09, rana16}).  We  suspect that only the final two epochs in \citet{han92} might  represent the source  in quiescence (taken on 1991 January 31 and 1991 May 31), but we cannot be certain.   We therefore conservatively exclude all epochs already reported by \citet{han92} from our study, and we begin our dataset  with a VLA epoch taken on 1991 September 25.   Although, we stress that none of our results are (qualitatively) affected if we were to include the final two epochs from \citet{han92}.  

We found  a total of \nvlaWithBad\ observations over five  observing frequencies from 1.4 - 22.5~GHz taken with the historical VLA, i.e., before the  VLA was upgraded and re-dedicated as the Karl G.\ Jansky VLA in 2012.  The majority of these data are  at 8.4 GHz (\nvlaXwithBad\ observations).  We also found five observations with the upgraded VLA from 4--8 GHz and four taken from 8--12 GHz.  We also include \nvlba\ observations with the Very Long Baseline Array (VLBA) at 5.0 GHz  (\nvlbac\ observations in 2014)  and at 8.4 GHz (\nvlbax\ observation in 2008).   A summary  of observations is provided in Table~\ref{tab:nobs}, and a catalog of flux densities  in Table~\ref{tab:obslog}.  Note that there are two  observations at 8.4 GHz for which  we could not obtain good calibration solutions.  We include entries for those observations in our catalog (Table~\ref{tab:obslog})  for completeness, but we have omitted them from the tally of  observations in Table~\ref{tab:nobs}.  In total, we measure flux densities (or limits) for \nallobs\ observations.  

\begin{deluxetable}{c c c c}
\tablecaption{Number of Observations per Frequency  \label{tab:nobs}}
\decimals
\tablecolumns{4}
\tabletypesize{\footnotesize}
\tablewidth{8in}
\tablehead{
                \colhead{Observing Band }                               & 
                \colhead{Telescope}                       & 
                \colhead{$N_{\rm obs}$}                               & 
                \colhead{$N_{\rm det}$}            
}
\colnumbers          
\startdata     
L /1.4 GHz                         &    Historical VLA      &    4       &    1      \\ \hline
C / 4.9 GHz                        &   Historical VLA       &    24      &    14     \\ 
\nodata                &       Upgraded VLA  &      5\tablenotemark{a}      &     5     \\  
\nodata                &        VLBA              &  13     &     13     \\ \hline
X / 8.4 GHz                         &    Historical VLA       &   84\tablenotemark{b}       &   59     \\
\nodata                &     Upgraded VLA    &    4\tablenotemark{a}        &   4      \\  
\nodata                &      VLBA                 &     1        &   1       \\ \hline
K$_{\rm U}$ / 14.9 GHz         &   Historical VLA      &      11      &    5     \\ \hline
K / 22.5 GHz                         &   Historical VLA      &        4       &    0      \\
 \enddata
\vspace{0.3cm}
\tablenotetext{a}{Four observations with the upgraded VLA post-2015 outburst were taken in subarray mode, with half of the antennas observing at C-band and the other half at X-band, which we count as separate C- and X-band observations in this table.  The fifth upgraded VLA observation was taken at C-band (4-8 GHz) in 2013.  We often add a 7.7 GHz flux density measurement from this C-band observation into our analysis of the X-band sample (which is not reflected within this table).}
\tablenotetext{b}{There are two additional observations in the archive at this frequency for which we could not obtain a calibration solution.}
\tablecomments{Column (1) observing band and frequency. 
Column (2) the telescope. 
Column (3) the number of observations.
Column (4) the number of detections.}
\end{deluxetable}
\renewcommand\arraystretch{1}

\begin{deluxetable*}{c c c C C C C C c c c }
\tablecaption{Catalog of Radio Observations \label{tab:obslog}}
\decimals
\tablecolumns{11}
\tabletypesize{\footnotesize}
\tablehead{
                \colhead{Date}                               & 
                \colhead{MJD}                        & 
                \colhead{Program ID}                               & 
                \colhead{Configuration}                               & 
                \colhead{$\tau_{\rm source}$}             & 
                \colhead{Frequency}                     & 
                \colhead{$f_\nu$}                   & 
                \colhead{$\alpha_r$}    &   
                \colhead{Primary}  &   
                \colhead{Secondary} &  
                \colhead{PI}     \\
                \colhead{}                                      & 
                \colhead{}                                       & 
                \colhead{}                                      & 
                \colhead{}                                        & 
                \colhead{(min)}                        & 
                \colhead{(GHz)}                                       & 
                \colhead{(mJy)}               &    
                \colhead{}                         &  
                \colhead{Calibrator}         & 
                \colhead{Calibrator}    &   
               \colhead{}    
}
\colnumbers
\startdata     
1991 Sep 25     &    48525.012 & AH424        & {\rm BnA}                &         36.8 & 4.9          & 0.238 \pm 0.042          & 1.59 \pm 0.36            & 3C286           & 2025+337        & Han             \\
\nodata         &    48525.012 & AH424        & {\rm BnA}                &         38.7 & 8.4          & 0.560 \pm 0.046          & \nodata                  & 3C286           & 2025+337        & Han             \\
\nodata         &    48524.984 & AH424        & {\rm BnA}                &         55.7 & 14.9         & <0.695                   & \nodata                  & 3C286           & 2025+337        & Han             \\
1991 Oct 31     &    48561.022 & AH390        & {\rm BnA}                &         59.8 & 4.9          & 0.321 \pm 0.028          & -0.72 \pm 0.33           & \nodata         & 2025+337        & Hjellming       \\
\nodata         &    48561.004 & AH390        & {\rm BnA}                &         46.0 & 8.4          & 0.218 \pm 0.034          & \nodata                  & \nodata         & 2025+337        & Hjellming       \\
 \enddata
\tablecomments{This table is available in its entirety in the online journal.  Only a portion is shown here to illustrate its form and content.  
 Column (1) calendar date of each observation.  
Column (2) modified Julian day.
Column (3) program code for the VLA or VLBA.
Column (4) VLA configuration (or `VLBA' to denote VLBA observations). 
Column (5) dwell time on \src\ in minutes.
Column (6) observing frequency.  Historical VLA observations include 100 MHz of bandwidth, observations after the upgrade include up to 1024 MHz. 
Column (7) peak radio flux density at the  observing frequency in column (6).   All  error bars are reported at the 68\% confidence level (and they include systematic errors on the flux density calibration scale); upper limits are at the 5$\sigma_{\rm rms}$ level.  Blank entries indicate that we could not calibrate the observations.
Column (8) radio spectral index ($f_\nu \propto \nu^{\alpha_r}$) for epochs with multifrequency data taken within 30 min. 
Column (9) the primary flux calibrator used for each observation.  For blank entries, we manually set the flux scale using the expected flux density of the secondary calibrator, based on time-adjacent observations that used the same secondary calibrator at the same frequency.
Column (10) the secondary flux calibrator.
Column (11) principal investigator of observing program. }
\end{deluxetable*}

\subsection{Historical VLA}
\label{sec:histvla}
The vast majority of  data from the historical VLA were obtained  through  campaigns led by either Robert Hjellming or Michael Rupen.  All observations were taken in continuum mode, using two 50 MHz wide spectral windows.  We reduced historical VLA observations following standard procedures within the Astronomical Image Processing System version 31DEC14 \citep[{\sc aips};][]{greisen03}.  We  set the flux scale (using either 3C 48 or 3C 286) with the task {\sc setjy} and (time-dependent)  \citet{perley13} coefficients.   The complex gain solutions were then solved using scans of a secondary point-source calibrator (usually \phcal, although see Table~\ref{tab:obslog} for exceptions), and the flux scale was bootstrapped from the primary calibrator  using the task {\sc getjy}.     Primary flux calibrator scans were not included for 24 observations.  For those epochs,  we manually set the flux scale to the expected value of the secondary phase calibrator by interpolating the flux densities reported by {\sc getjy} of  time-adjacent observations of the same calibrator at the same frequency.  In these cases, we add a systematic uncertainty based on the level of variability of the secondary calibrator in nearby epochs, typically $\approx$5\%.   Finally, we add 5\% and 10\% systematic uncertainties to the flux scale for observations below and above 10 GHz, respectively.   

The data were  imaged using the task {\sc imagr}, using Briggs weighting with a robust value of zero to help minimise sidelobes from nearby sources in the field.    Of particular note is that a bright Jansky-level source lies 16.6 arcmin southeast of \src\ (\phcal, which was often used as the phase calibrator).   This source is difficult to deconvolve during the imaging process because of bandwidth smearing (given the 50 MHz spectral windows of the historical VLA). 
We visually inspected a random subset of images of \src\ over a range of  frequencies and array configurations.  We find that sidelobe artifacts from \phcal\  do not reach \src\ at observing frequencies $>$8 GHz. 
At  frequencies $<$8 GHz, 
there are multiple instances where artifacts from \phcal\ appear to increase the noise level near the location of \src, but there are not cases where those artifacts obviously bias the measured flux densities.

Flux densities were measured using the task {\sc jmfit}, by fitting a two-dimensional Gaussian (fixed to the width of the synthesized beam) at the known location of \src.   Given potential systematics raised by \phcal\ at lower frequencies, we  require detections to display peak flux densities at $>$5$\sigma_{\rm rms}$, where $\sigma_{\rm rms}$ is the root-mean-square (rms) noise measured in a blank region of the sky (with upper limits for non-detections calculated as $5\sigma_{\rm rms}$).

\subsection{Karl G. Jansky Very Large Array}
\label{sec:obs:jvla}
Our sample also includes five epochs with the upgraded VLA. The first was taken in 2013 \citep{rana16}, and the other four were taken at the  end of the 2015 outburst after the system  re-entered quiescence  \citep{plotkin17}.  

\subsubsection{2013 Quiescence}
\label{sec:obs:rana}
The 2013 observation lasted $\approx$9  hours in C band (4-8 GHz) in B configuration (maximum baseline\,$\approx 11$\, km), under program code 13B-016 (PI Corbel).   Two basebands of bandwidth 1024 MHz were  placed at 5.25 and 7.45 GHz. The sources 3C 286 and \phcal\ were used as the primary and secondary calibrators, respectively.  

Similar data reduction steps were performed as described in Section \ref{sec:histvla},  except for the following: we used the  Common Astronomy Software Application v5.1.1 \citep[{\sc casa};][]{mcmullin07}, to allow us to account for the larger fractional bandwidth; we Hanning smoothed the data to avoid radio frequency interference from bleeding into nearby frequency channels; we  used the primary flux calibrator  to solve for  delay and complex bandpass solutions; and we  imaged the field using the task {\sc clean}, using Briggs weighting  (robust=1.0), and two Taylor terms to model the frequency dependence over the 2 GHz bandwidth.  In these observations, the effects  of bandwidth smearing on \phcal\ were not significant, on account of the smaller frequency channels (2 MHz vs.\ 50 MHz), so we were able to adequately deconvolve \phcal.   Flux densities were measured by fitting a two-dimensional Gaussian with the {\sc casa} task {\sc imfit}, forcing a point source (fixed to the width of the synthesized beam) at the known location of \src.

\subsubsection{2015 Post-outburst}
\label{sec:obs:outburst}
Four observations  from the end of the 2015 decay are included in this study (from 28 July - 5 August), under program code SG0196 (PI Plotkin).  As noted at the beginning of Section~\ref{sec:obs}, \src\  re-entered  quiescence by the time these observations were taken, according to its X-ray spectral behavior \citep{plotkin17}.  All four observations were taken in the most extended A configuration (maximum baseline $\approx$30 km) using the VLA in subarray mode, where about half of the antennas observed from 4-8 GHz, and the other half observed from 8-12 GHz, yielding strictly simultaneous multi-frequency coverage over four frequencies centered at 5.2, 7.5, 8.6, 11.0 GHz (with 1024 MHz bandwidth at each frequency).   These data were reduced using the same procedures as described in Section~\ref{sec:obs:rana} (see \citealt{plotkin17} for details).

\subsection{VLBA}
We monitored \src\ with the VLBA over 13 (approximately) fortnightly observations between 2014 February 3 and August 22, under program code BM399 (PI Miller-Jones).   Each observation lasted 2\,hr, yielding $\approx$56\,min on source.  We observed with all available antennas, using 256 MHz of bandwidth centered on a frequency of 4.98\,GHz.   We used J2025+337 \citep[][16.6\,arcmin from \src]{ma98} as both the fringe finder and phase reference calibrator, and the somewhat more distant J2023+3153 \citep[][1.99$^{\circ}$ from \src]{ma98} as an astrometric check source.    We also identfied an additional VLBA observation at 8.4 GHz reported by \citet{miller-jones09} taken on 2008 November 17 under program code BM290, taken in dual circular polarization with 64 MHz bandwidth per polarization (133 min on source), which we re-reduced.\footnote{Program BM290 contained four other observations at 8.4 GHz, which were already included in our sample because they used the phased VLA with the VLBA.  We used the phased VLA observations in preference to the VLBA, because the phased VLA uses a standard flux calibrator, while the VLBA observations rely on system temperatures for amplitude calibration.} 

We calibrated the data using {\tt AIPS} (version 31DEC15) following  standard procedures.  For the 8.4 GHz observation, we used geodetic blocks at the start and end of the observation to correct for unmodelled clock errors and tropospheric delays.  For all observations, we applied updated Earth orientation parameters, and we corrected for ionospheric dispersive delays using total electron content maps.  We used system temperature information for amplitude calibration, corrected the phases for parallactic angle effects as the antenna feeds rotated with respect to the sky, and iteratively imaged and self-calibrated the phase reference source J2025+337 to derive a model for calibrating the phase, delay and rate solutions, which were then applied to the target.  V404 Cygni was too weak for self-calibration.  After imaging with natural weighting (for maximum sensitivity), we determined the source flux density by fitting a point source in the image plane.  \src\ remains point-like at VLBA resolutions in quiescence \citep{miller-jones08}, such that we do not expect to resolve out any jetted radio emission, allowing a fair flux density comparison between VLA and VLBA epochs.

\begin{figure*}
\begin{center}
\includegraphics[scale=0.48]{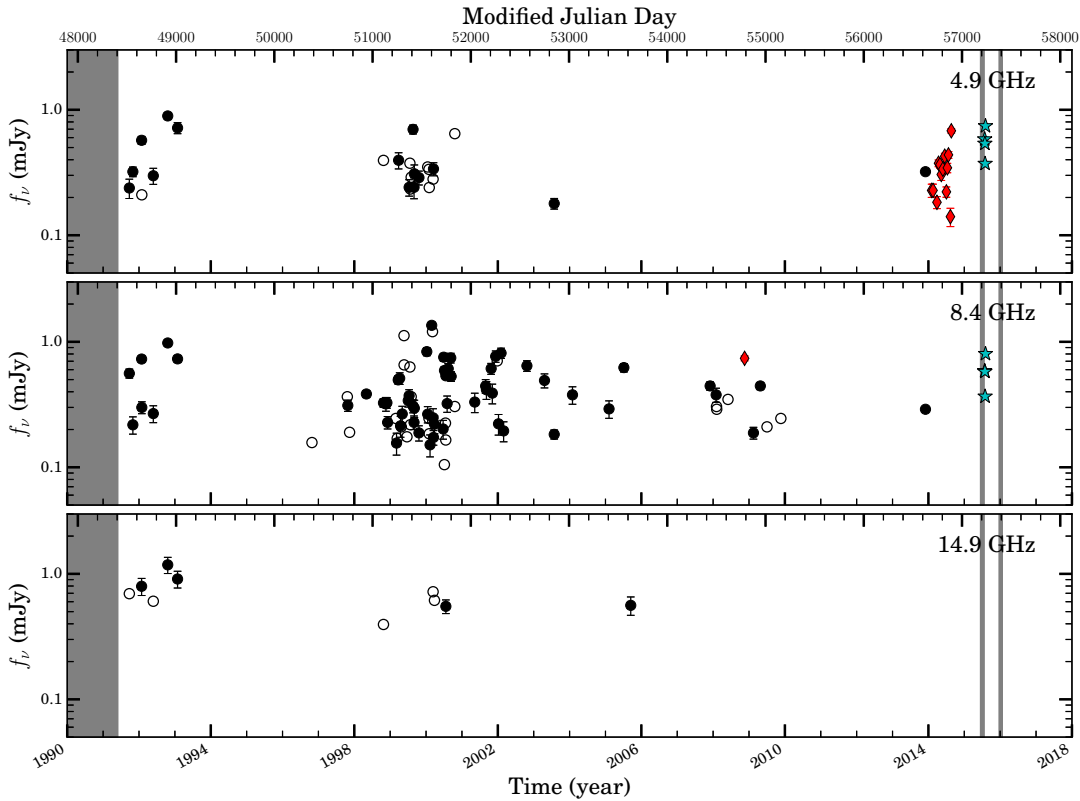}
\caption{Radio light curves spanning 1991-2015 from the VLA and VLBA (the numbers of observations at each frequency are summarised in Table~\ref{tab:nobs}).  We only show frequencies here with at least five data points.  Filled symbols represent detections and open symbols represent $5\sigma_{\rm rms}$ upper limits.  The red diamonds represent epochs taken with the VLBA, and the cyan stars represent four epochs taken between July and August of 2015 at the end of the outburst, after \src\ had re-entered quiescence.  Error bars are  often smaller than the size of each symbol.  The grey shaded regions mark time periods during the 1989 and 2015 X-ray outbursts, and the 2015-2016 mini-outburst.  }
\label{fig:lc}
\end{center}
\end{figure*}
 
\section{Results}
\label{sec:res}

\subsection{Long-term  Variability}
\label{sec:res:var:long}
\begin{figure}
\includegraphics[scale=0.4]{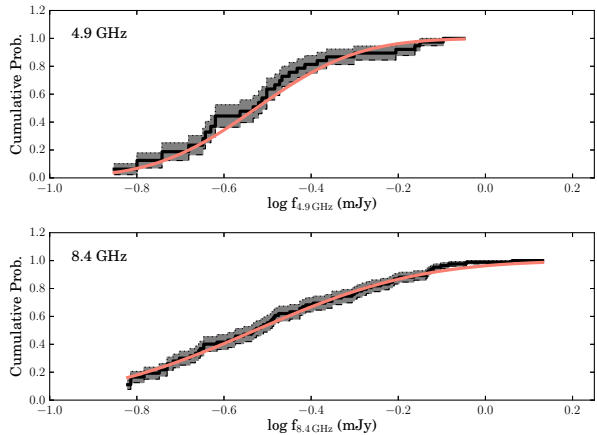}
\caption{The cumulative distribution of the logarithm of flux densities (calculated as $1-S(\log f_\nu)$, where $S(\log f_\nu)$ is the survival function  from the Kaplan-Meier estimator).  The gray shaded area represents the 90\% confidence interval.    The top panel is for 4.9 GHz (\nstatscband\ observations), and the bottom panel is 8.4 GHz (\nstatsxband\ observations). 
Both flux density distributions are consistent with lognormal distributions, which are illustrated by red solid lines, where $\left<\log f_\nu\right> = -0.53 \pm 0.19$ and $-0.53 \pm 0.30$ at 4.9 and 8.4 GHz, respectively, where $f_\nu$ is the flux density in mJy, and errors represent standard deviations of the lognormal distributions.}
\label{fig:fluxhist}
\end{figure}

Light curves are shown in Figure~\ref{fig:lc} at 4.9, 8.4, and 14.9 GHz (we omit our other two frequencies, 1.4 and 22.5 GHz as they each contain only 4 observations).    To quantify the distribution of flux densities at each frequency in the presence of non-detections, we perform a survival analysis.  Using {\tt survfit} in {\tt R} (within the {\tt survival} package\footnote{\url{https://CRAN.R-project.org/package=survival}}
) we calculate the survival function, $S(\log f_\nu)$, via the Kaplan-Meier estimator \citep[see, e.g.,][for a description of the Kaplan-Meier estimator and  examples of astrophysical applications]{feigelson85}.  We then estimate the cumulative distribution function as $P(\log f_{\nu}) = 1 - S(\log f_\nu)$, which we display in Figure~\ref{fig:fluxhist} at 4.9 and 8.4 GHz (omitting the four observations from 2015).  In Section~\ref{sec:obs:spind} we image the 2013 observation with the upgraded VLA at four separate frequencies, centered at 5.0, 5.5, 7.2, and 7.7 GHz (each using 512 MHz bandwidth), to measure a spectral index.  We therefore include the 5.0 GHz flux density from 2013 in the 4.9 GHz distribution here, and the 7.7 GHz flux density  in the 8.4 GHz distribution.    In total, our 4.9 and 8.4 GHz distributions include \nstatscband\ and \nstatsxband\ data points, respectively (of which \nstatscbandLim\ and \nstatsxbandLim\ are upper limits, respectively).  Detections (at the $>$5$\sigma_{\rm rms}$ level)
 range between 0.14 -- 1.35 mJy bm$^{-1}$.

We compare the cumulative distribution functions in Figure~\ref{fig:fluxhist}  to lognormal distributions using the Peto \& Peto modification of the Gehan-Wilcoxon test, as implemented by {\tt cendiff} in the {\tt R} package {\tt Nondetects and Data Analysis for Environmental Data (NADA\footnote{\url{https://CRAN.R-project.org/package=NADA}}}
).  The distributions of flux densities from \src\ are not statistically different from lognormal distributions ($p=0.81$ and 0.93  at 4.9 and 8.4 GHz, respectively) with $\left<\log f_{\rm 4.9}/{\rm mJy}\right> = -0.53\pm 0.19$ and $\left< \log f_{\rm 8.4}/{\rm mJy}\right> = -0.53\pm 0.30$, where $f_{\rm 4.9}$ and $f_{\rm 8.4}$ are flux densities at 4.9 and 8.4 GHz, respectively, in units of mJy.    The quoted errors represent standard deviations on the lognormal distribution (i.e., they are not  errors on the mean).\footnote{The quoted numbers for the lognormal distributions are not biased by a small number of individual observations in the small- or large flux density tails of the distributions.  We show this by bootstrapping the flux density distributions 100 times at each frequency, selecting subsamples of 30 and 50 at 4.9 and  8.4 GHz, respectively, with replacement.  After running the same survival analysis on each bootstrapped distribution, we find  average values from the 100 distributions to be $\left<\log f_{\rm 4.9}/{\rm mJy}\right> = -0.54\pm 0.19$ and $\left< \log f_{\rm 8.4}/{\rm mJy}\right> = -0.54\pm 0.28, where errors represent  standard deviations$}  

We find that \src\ shows significant flux density variations that are in excess of statistical fluctuations from measurement errors (which are typically $\pm$0.04 mJy bm$^{-1}$ with the historical VLA).  Taking 8.4 GHz as example, among  \nstatsxbandDet\ detections we find a reduced $\chi^2_r$ of 46 (for \nstatsxbandDof\ degrees of freedom, as compared to a model with constant flux density).  We note that $\chi^2_r$ is slightly biased by one observation from 2013 with higher signal-to-noise using the upgraded VLA.    Removing that observation still suggests significant intrinsic variability ($\chi_r^2 = 33$ for \nstatsxbandDofVlaonly\ degrees of freedom). 
The fractional rms variability for all \nstatsxbandDet\ observations at 8.4 GHz  is $F_{\rm var} = 54 \pm 6$ \% \citep{vaughan03}, which   is not biased by the  higher signal-to-noise observations (i.e., we also calculate $F_{\rm var} = 54 \pm 6\%$ if we exclude the 2013  observation).  If we consider the statistical fluctuations induced by variability to have a standard deviation of $\sigma_{f_\nu, \rm var} = \pm f_\nu  F_{\rm var}$, then propagating errors would yield $\sigma_{\log f_\nu, {\rm var}} = \pm F_{\rm var}/\ln 10$, such that a 54\% fractional rms variability translates to 0.23 dex in logarithmic  space.

\begin{figure}
\includegraphics[scale=0.45]{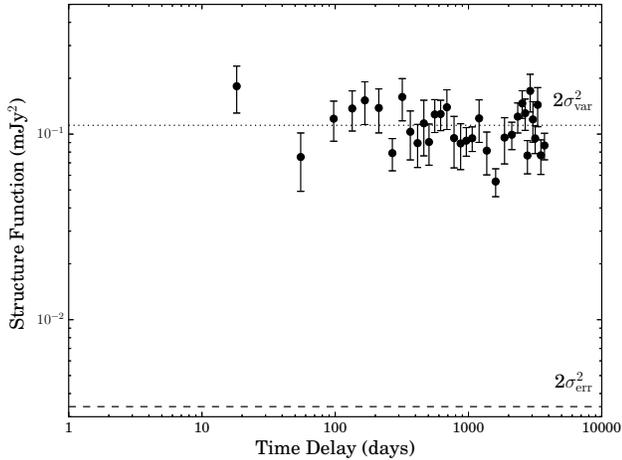}
\caption{The first-order structure function, including observations taken between 7.7 -- 8.4 GHz (\nstatsxbandDet\ measurements, omitting upper limits).    We bin the structure function to contain 50 data points per time delay.  Horizontal dashed and dotted lines illustrate, respectively, twice  the average measurement error squared (2$\sigma_{\rm err}^2$) and twice the variance (2$\sigma_{\rm var}^2$) of the \nstatsxbandDet\ data points (note, the dotted line is not a fit).    Flux density variations are well in excess of statistical fluctuations from measurement errors.  The flat slope indicates either flicker noise variations (i.e., a power function $P(F) \propto F^{-1}$), or shot noise disturbances that resemble (uncorrelated) white noise when probed on timescales longer than a characteristic damping timescale of $\lesssim$10 days.}
\label{fig:sf}
\end{figure}

To further quantify the flux variability we produce the first-order structure function $V(\tau)$, which characterizes the amount of variability power as a function of time scale in the case of  irregularly sampled data, 
\begin{equation}
V(\tau) = \left< \left[f \left(t + \tau \right) - f \left(t \right)\right]^2\right>,
\end{equation}
where $f(t)$ is a flux density measurement at time $t$, and $\tau$ is a time delay.    In the following analysis, we include only the \nstatsxbandDet\ detections at 8.4 GHz.  For every  pair of data points ($t_i, t_j$) in our light curve, we calculate $V_{ij}(\tau_{ij}) = \left[f \left(t_j\right)) - f \left(t_i\right)\right]^2$, where $\tau_{\rm ij} = t_j - t_i$.  We then bin our set of $V_{\rm ij}$ measures by time difference ($\tau_{\rm ij}$) so that each bin contains 50 data points, and we take the average of those 50 measurements to calculate $V(\tau)$ (where we adopt the midpoint of all time differences within each bin as the value for the time delay $\tau$).\footnote{Our error bars represent the error on the mean value of $V(\tau)$ in each time bin.  We define our errors in this manner  for ease of comparison to \citet{bernardini14} who provide a structure function for the quiescent X-ray variability of \src.}    The structure function is shown in Figure~\ref{fig:sf}, where we probe long-term variability over timescales of $\sim$10 -- 4000 days. 
   
   The slope of the structure function provides information on the power distribution of flux variations (see, e.g., \citealt{hughes92} for details, which we summarise below).  For example, if variability is characterised as flicker noise (i.e., a power function $P(F) \propto F^{-1}$, where $F$ is the inverse timescale, i.e., frequency, of fluctuations), then the structure function $V(\tau)$\,=\,constant.  The constant is expected to be 2$\sigma_{\rm var}^2$, with $\sigma_{\rm var}^2$ being the variance of the observed flux densities.  One could also obtain $V(\tau)=2 \sigma_{\rm var}^2$ on long timescales if the structure function probes white noise  (e.g., one interpretation would be that the probed timescales   are longer than the characteristic  timescale on which shot noise variations dampen).  As another example,  red noise variations ($P(F) \propto F^{-2}$) produce first-order structure functions of the form $V(\tau) \propto \tau$, which can often be interpreted as disturbances that follow a random-walk process \citep[e.g.,][]{kelly09}.  Finally, flux variations smaller  than the level of a typical error bar on flux density measurements ($\sigma_{\rm err}$) are not  meaningful (assuming that $\sigma_{\rm err}$ is dominated by statistical noise).  Thus, one expects the structure function to satisfy $V(\tau) \geq 2 \sigma_{\rm err}^2$ at all timescales.  
        
    The structure function in Figure~\ref{fig:sf} appears flat, with a best-fit slope and normalisation of  $\beta=-0.11\pm0.07$ and $V_0 = 0.16\pm0.04$  mJy$^{2}$ (where $V(\tau) = V_0 \tau^\beta$).   Measuring the variance directly from the \nstatsxbandDet\ data points yields 2$\sigma_{\rm var}^2 = 0.11$ mJy$^{2}$.  Therefore,  the structure function appears consistent with plateauing  near 2$\sigma_{\rm var}^2$, which signifies either flicker or white noise.  In the latter case, we would be probing jet disturbances on timescales ($\gtrsim$10 days) that are longer than  characteristic damping timescales.  
    
\begin{figure*}
\includegraphics[scale=0.44]{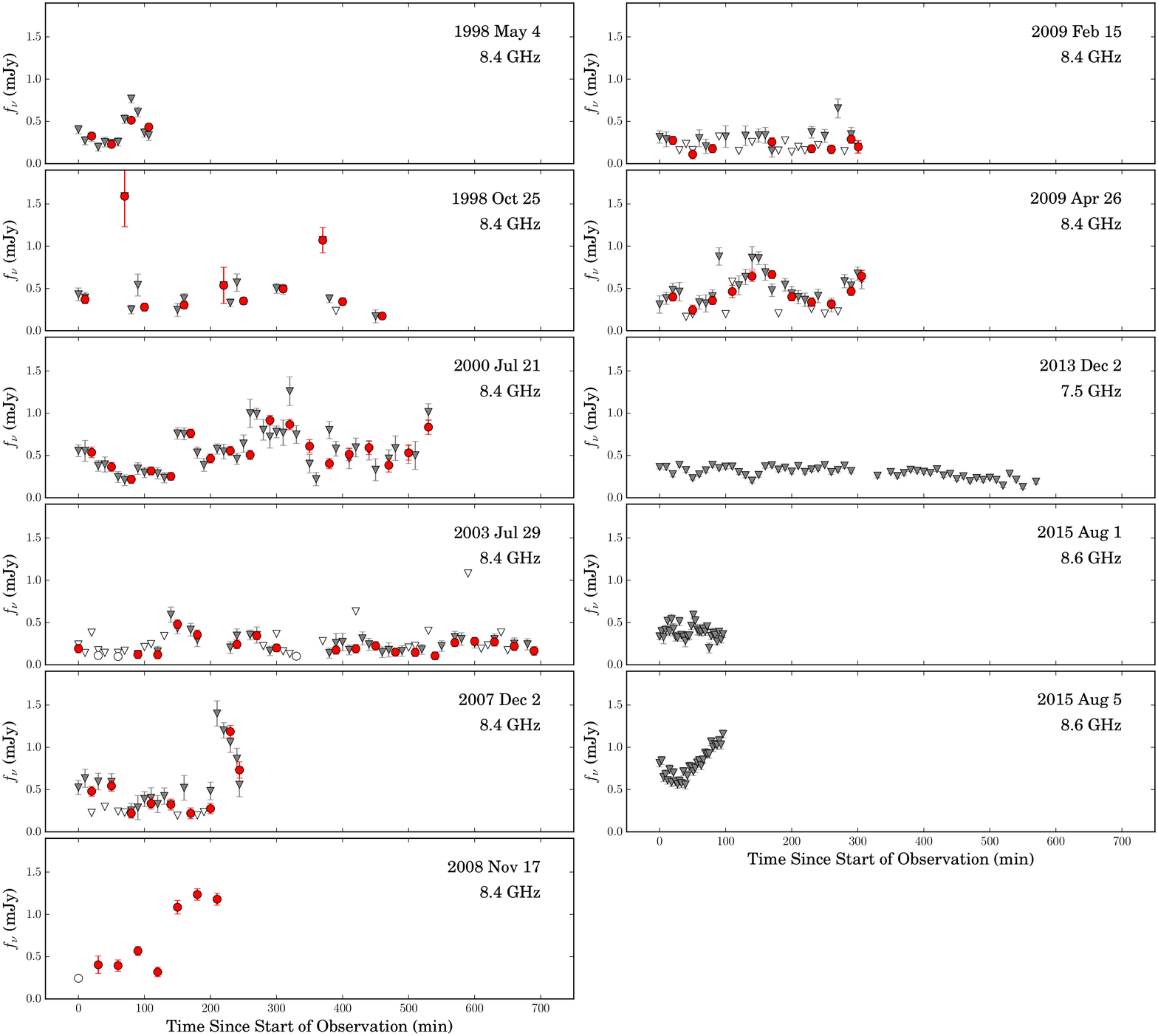}
\caption{Eleven observations that are long enough to produce light curves on sub-hour time resolution. Grey triangles represent 10 min time bins (3 min for the final two observations in 2015), and open triangles are 2$\sigma_{\rm rms}$ upper limits.  For the historical VLA (observations from 1998-2009) we overplot 30 min time bins as red circles (2$\sigma_{\rm rms}$ limits as open circles), and for the VLBA observation (2008 November 17) we only include 30 min time bins.  Note the slightly different central frequencies for each light curve (top right of each panel).  Flares that increase the flux density by factors of 2-4 are common on minute to hour timescales, but there is not a single template that can describe all flares in terms of their  amplitudes, rise times, and decay times. Note that the three epochs from 2015 were taken at the very end of the 2015 outburst decay, after the source had re-entered quiescence according to its X-ray signatures \citep[see][]{plotkin17}.} 
\label{fig:shrtlc}
\end{figure*}

\subsection{Short-term Variability}
\label{sec:res:var:short}

We also examine  variability on sub-hour timescales, focusing on  observations long enough (usually $\gtrsim$90 min on source) to produce light curves over multiple time bins (we are able to achieve time resolutions ranging from 3-30 min, see Figure~\ref{fig:shrtlc}).   We  focus  on observations near 8 GHz where we have 11 long observations total, including  seven from the historical VLA, one  from the VLBA, and three from the upgraded VLA.

\begin{deluxetable}{c c c C}
\tablecaption{Short-term Variability Statistics \label{tab:varstats}}
\decimals
\tablecolumns{4}
\tabletypesize{\footnotesize}
\tablewidth{8in}
\tablehead{
                \colhead{Date}                               & 
                \colhead{$\left(\chi^2_r\right)_{\rm flux}$}                       & 
                \colhead{deg of freedom}                               & 
                \colhead{$\beta$}                 
}
\colnumbers          
\startdata     
1998 May 04     & 11.9            & 11              & \nodata         \\ 
1998 Oct 25     & 9.0             & 14               & \nodata         \\ 
2000 Jul 21     & 8.2             & 38               & \nodata         \\ 
2003 Jul 29     & 2.1             & 26             & \nodata         \\ 
2007 Dec 02     & 8.5             & 16              & \nodata         \\ 
2008 Nov 17     & 35.4            & 6                & \nodata         \\ 
2009 Feb 15     & 1.5             & 13              & \nodata         \\ 
2009 Apr 26     & 3.7             & 23             & \nodata         \\ 
2013 Dec 02     & 7.7             & 52            & 0.29 \pm 0.05   \\ 
2015 Aug 01     & 2.8             & 32              & 0.10 \pm 0.11   \\ 
2015 Aug 05     & 12.3            & 32            & 1.08 \pm 0.08   \\
 \enddata
\tablecomments{Column (1) calendar date of each observation. 
Columns (2)-(3) reduced $\chi^2_r$ for each light curve (including only detections) and the number of degrees of freedom.
Column (4) the best-fit powerlaw index to the structure function, when available ($V(\tau) \propto \tau^\beta$).  }
\end{deluxetable}
\renewcommand\arraystretch{1}

\begin{figure}
\includegraphics[scale=0.5]{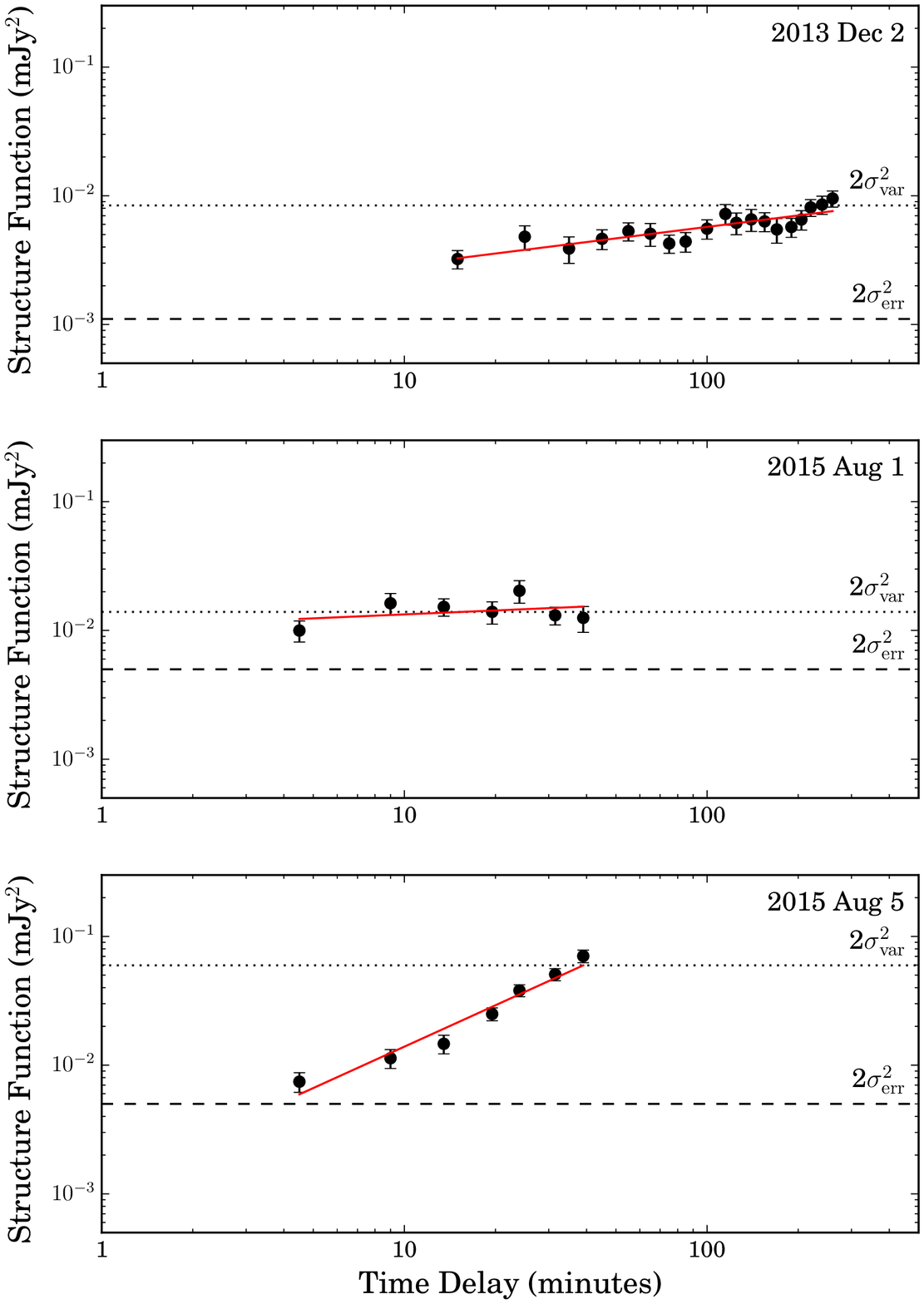}
\caption{First order structure functions for observations with the upgraded VLA, binned to 50 data points per time delay.   Symbols and lines have the same meaning as in Figure~\ref{fig:sf}), except that the red lines represent fits to the data.  Note that even when \src\ does not show obvious flares  (e.g., 2015 August 1), the observed flux density variations are still in excess of expectations from measurement errors.    The structure function on  2015 August 5 is consistent with red noise ($V(\tau) \propto \tau$). }
\label{fig:sfshrt}
\end{figure}

We characterize the level of variability on each epoch by calculating the reduced $\chi^2_r$ of the flux densities in each light curve.  For this calculation we ignore upper limits (which  causes us to underestimate $\chi^2_r$ in some cases).    Our measured $\chi^2_r$ values are listed in Table~\ref{tab:varstats}.  Treating $\chi^2_r > 3$ as a crude diagnostic for significant intrinsic variations, we find obvious variability on 8/11 epochs.  However, we cannot determine from $\chi^2_r$ alone if we should consider the  other three epochs as non-variable.  Rather, lower $\chi^2_r$ values imply that variability on those epochs is less extreme relative to fluctuations from statistical noise, and  further tests are required (see next paragraph).  

 The three epochs with the upgraded VLA each display different short-term variability characteristics (i.e., the flux density is slowly decreasing with time on 2013 December 2, there are no obvious flares on $>$3 min timescales on 2015 August 1, and we observe the beginning of a flare in 2015 August 5).  These three epochs therefore provide useful illustrative examples, and as such we display their structure functions in Figure~\ref{fig:sfshrt}.  We fit a powerlaw to each structure function, and we find powerlaw indices of $\beta= 0.29 \pm 0.05, 0.10 \pm 0.11$ and $1.08\pm0.08$ on 2013 December 2, 2015 August 1, and 2015 August 5, respectively ($V(\tau) \propto \tau^\beta$).    Note that the structure function on 2015 August 1 is flat, plateauing at $2 \sigma_{\rm var}^2$, indicating that \src\  indeed shows (uncorrelated) variability in excess of statistical measurement noise fluctuations on 2015 August 1, even though $\chi_r^2 = 2.8$.       

\subsubsection{Quiescent Flares}
\label{sec:res:qflares}

In this subsection we summarise some of the flaring behaviour of \src\ in quiescence.  We note that factor of $>$2 changes in flux density appear to be common.  However, we do not attempt to define a duty cycle, on account of the limited number of observations (11) with short-term light curves, for which we do not have uniform time resolution.

The 2007 December 2  observation \citep[first reported by][]{miller-jones08} provides an  extreme example of rapid short-term variability, where the 8.4 GHz flux density increased by a factor of 3--4 in $<$10 min, reaching 1.4 mJy, followed by a slower decay that lasted at least 30 min.   (The beginning of that light curve also shows a series of three alternating detections and  non-detections,  suggesting a factor of $>$2.5 flux variability over 10 min).  However,  not all short-term variability follows a pattern of a fast rise followed by a slower decay.    For example, on 1998 May 4 a factor of 2-3 flare rose over $\approx$20 min, on  2009 April 26   a rise and a decay of a factor $\approx$3 in flux density occurred over $\sim$60-90 min in each direction, and on 2015 August 5  \src\ displayed an increase in flux density by a factor $>$2 over $>$60 min.

 We also see a potential variety in decay timescales.  For example, the 2007 December 2 flare that quickly rose in $<$5-10 min took at least 30 min to decay.  At the other extreme,  the $\approx$9 hour 2013 December 2 observation appears to be decreasing in flux the entire time (with some smaller-scale variations superposed).  If a flare preceded the beginning of that observation, then that implies some flares decay on timescales of at least several hours.  From the above we conclude that tens of minutes to hours represent reasonable minimum characteristic timescales for the damping of radio flares.  Although flares appear common, \src\ also undergoes quieter periods of time, where either flares are absent or low-amplitude flares  occur on short timescales  $<$3 min (e.g., 2015 August 1).

\begin{figure*}
\includegraphics[scale=0.45]{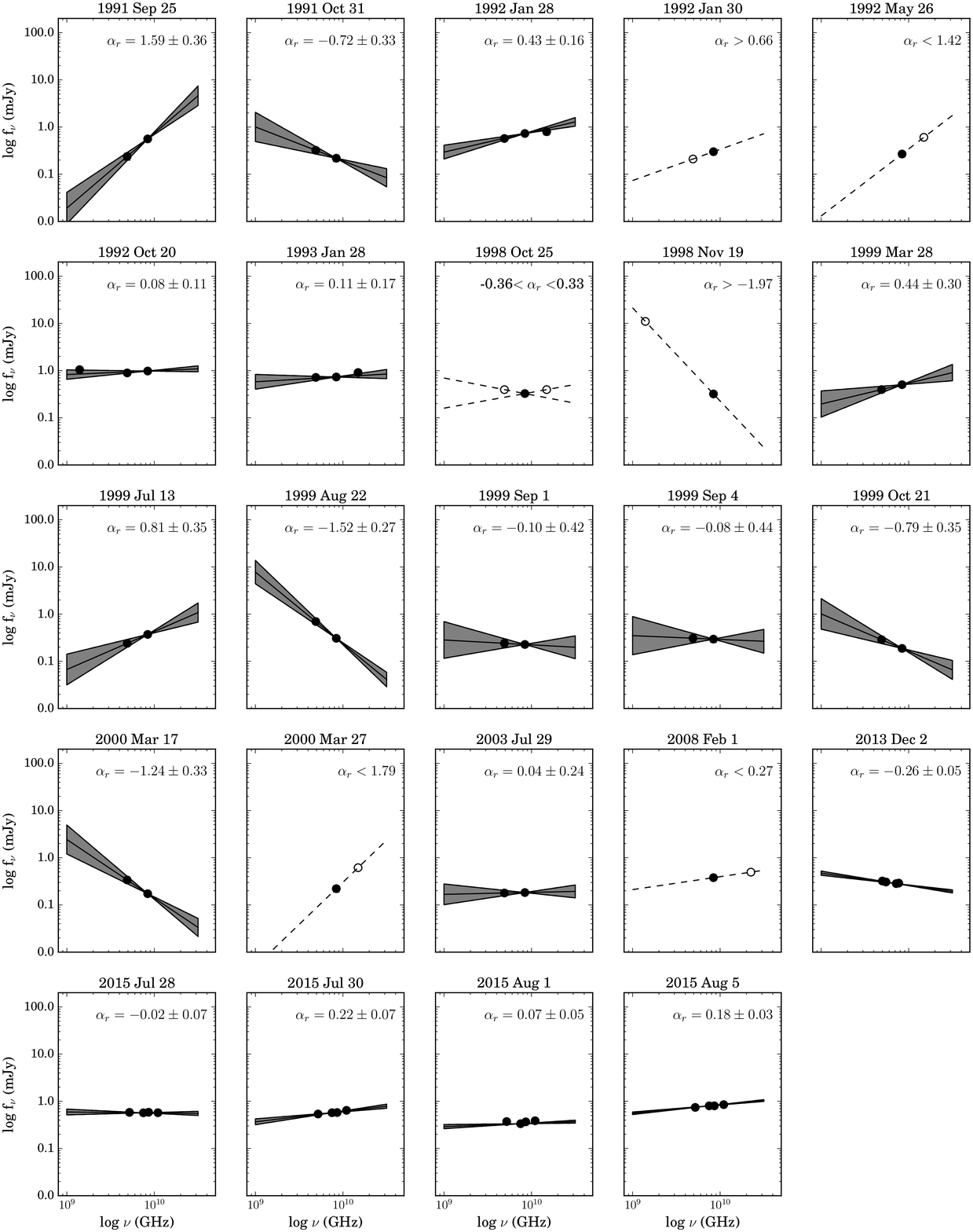}
\vspace{-1.7cm}
\caption{Epochs with multifrequency information taken within 30 min, with each panel showing detections as solid symbols (error bars are smaller than the size of each symbol) and 5$\sigma_{\rm rms}$ upper limits as open circles.  The radio spectral index  $\alpha_r$ (or limit) is provided in the top right corner of each panel, with the gray shaded regions representing 68\% confidence intervals on $\alpha_r$ (and dashed lines representing limits).   Note that we observe a range of  steep, flat, and inverted spectra, but non-simultaneity could be exaggerating the true spread in $\alpha_r$ (except for the final five epochs where the multifrequency information from the upgraded VLA is strictly simultaneous).  The average spectral index is consistent with a flat radio spectrum, $\left<\alpha_r\right> = 0.02 \pm 0.65$.   However, the 2013 December 2 observation highlights that the strictly simultaneous spectral index can be negative on some epochs ($\alpha_r = -0.26 \pm 0.05$). }
\label{fig:spinds}
\end{figure*}

\subsection{Radio Spectra}
\label{sec:obs:spind}
For 24 epochs  there are observations at multiple frequencies within $\pm$30 min, from which we measure spectral indices ($f_{\nu} \propto \nu^{\alpha_r}$) as shown in Figure~\ref{fig:spinds}.   For epochs with exactly two  frequencies, we calculate $\alpha_r$ (or place a limit) analytically as $\alpha_r = \ln(f_{\nu_1}/f_{\nu_2}) / (\nu_1/\nu_2)$, where  $f_{\nu_1}$ and $f_{\nu_2}$ refer to flux densities at frequencies $\nu_1$ and $\nu_2$, respectively.   We assign an error on $\alpha_r$ by propagating through  statistical errors (we ignore errors on frequency for the historical VLA given the relatively small bandwidth at each frequency, and we also ignore systematic errors from short-term variability since we generally do not know if the spectra were taken during flaring activity).   If $>$2 observing frequencies, then we measure $\alpha_r$ through a least squares fit to the spectrum (in log space).  We estimate error bars through Monte Carlo simulations where we randomly add  noise to each flux density measurement (assuming Gaussian noise with a standard deviation equal to the measurement error on each data point) and then refit the spectral index.  We repeat 1000 times and estimate $\sigma_{\alpha_r}$ as the standard deviation on the resulting $\alpha_r$ distribution.

For the 2013 epoch from \citet{rana16} with the upgraded VLA, we create four (strictly simultaneous) images by splitting the bandwidth into four sub-bands with 512 MHz bandwidth (centered at 5.0, 5.5, 7.2, and 7.7 GHz), and we  allow the frequency to randomly vary across each 512 MHz when running our Monte Carlo simulations to estimate error bars).   We obtain $\alpha_r=-0.26 \pm 0.05$, which is consistent with the value $\alpha_r = -0.27 \pm 0.03$ obtained by \citet[][]{rana16}. For the four VLA observations from 2015, we adopt the spectral indices reported by \citet{plotkin17}, which were calculated using the same least squares fitting method as described above (over 4--12 GHz). 

\subsubsection{Long-term Spectral Variations}
\label{sec:obs:spind:long}
\begin{figure}
\includegraphics[scale=0.45]{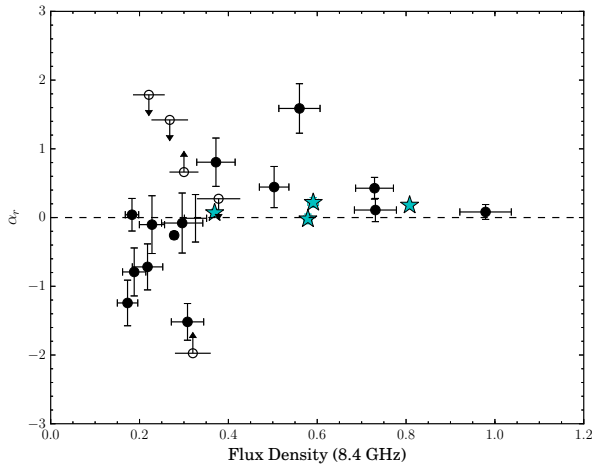}
\caption{Radio spectral index $\alpha_r$ versus 8.4 GHz flux density.  The  cyan stars represent the four epochs from 2015.  The five data points with the smallest error bars were taken with the upgraded VLA, and the other 19 data points were taken with the historical VLA.  The dashed horizontal line marks $\alpha_r=0$ for reference.  The spectral index tends to become negative only at low flux densities, which may be related to observational effects, such as non-simultaneity and/or  most low flux density observations having larger measurement errors.  However, some negative spectral indices at low flux densities are likely reflecting intrinsic changes within the jet (e.g., the  2013 December 2 epoch with a well-measured $\alpha_r = -0.26 \pm 0.05$ from strictly simultaneous multifrequency data).}
\label{fig:spindflux}
\end{figure}

We measure a large range of spectral indices, from $-1.5 \lesssim \alpha_r \lesssim +1.6$, with the spread of $\alpha_r$ being larger at lower flux densities (Figure~\ref{fig:spindflux}).  However, we argue in Section~\ref{sec:disc:spind} that this spread in $\alpha_r$ is likely  attributed largely to statistical and systematic errors (i.e., large error bars on most $\alpha_r$ measurements, especially at lower flux densities, combined with only five epochs having strictly simultaneous multifrequency data).

 To quantify the dispersion in radio spectral indices, we fit a Gaussian distribution to the  $\alpha_r$ measurements (using a Bayesian framework that allows for both upper and lower limits). We find a mean $\left<\alpha_r\right> = 0.02 \pm 0.17$ and a standard deviation $\sigma_{\alpha_r} = 0.65 \pm 0.15$, where the errors represent  68\% confidence intervals of the posterior distributions.   If we exclude the five spectra obtained from the upgraded VLA (which have significantly smaller error bars on $\alpha_r$), we find consistent results: $\left<\alpha_r\right> = -0.07 \pm 0.25$ and $\sigma_{\alpha_r} = 0.78 \pm 0.21$.    Throughout we adopt  $\left<\alpha_r \right>=  0.02 \pm 0.65$, which is consistent with a flat radio spectral index on average, as expected for a partially self-absorbed compact synchrotron jet.

\subsubsection{Short-term Spectral Variations}
\label{sec:obs:spind:short}

Among the 11  observations from which we produce sub-hour light curves in Section~\ref{sec:res:var:short}, four contain multi-frequency data allowing us to explore variations in $\alpha_r$ over sub-hour timescales.    These epochs include  2013 December 2 with the upgraded VLA  \citep[][4--8 GHz]{rana16},  two  epochs   at the end of the 2015 outburst on  August 1 and 5 \citep[4--12 GHz]{plotkin17}, and a  historical VLA observation  on 2003 July 29 that interleaved observations at 4.9 and 8.4 GHz every $\approx$15 min \citep{hynes04, hynes09}.  

\begin{figure*}
\begin{center}
\includegraphics[scale=0.45]{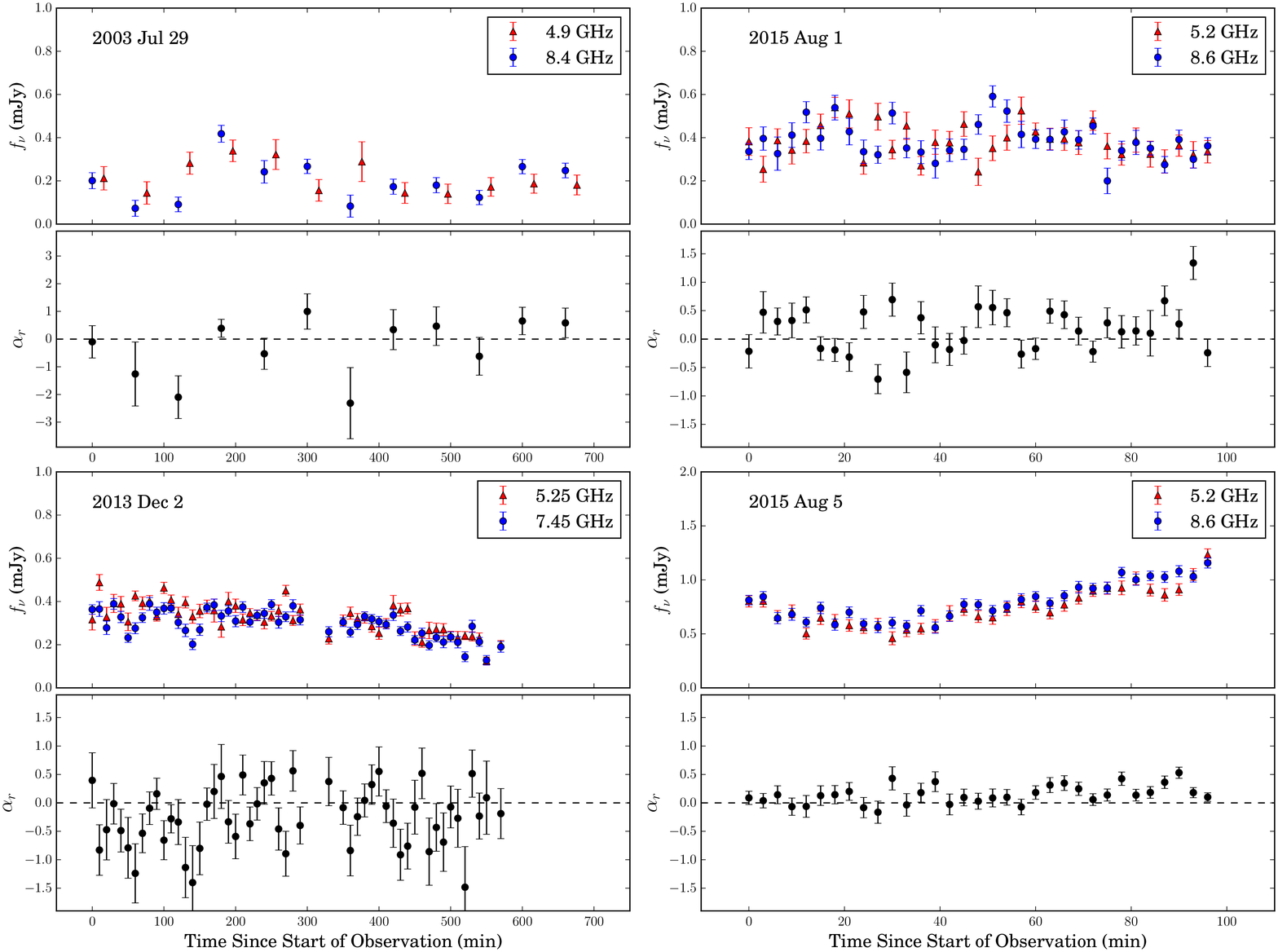}
\caption{Four observations for which we can extract spectral information over sub-hour  timescales, including 2003 July 29 (60 min time bins, also see \citealt{hynes09}), 2013 December 2 (10 min time bins, also see \citealt{rana16}), and 2015 August 1 and 2015 August 5 (3 min time bins, also see \citealt{plotkin17}). Note that the multi-frequency information is strictly simultaneous for the final three observations, but not for the 2003 observation.  We do not observe  meaningful fluctuations in $\alpha_r$ on these short time scales.}
\label{fig:shrtlcSpind}
\end{center}
\end{figure*}

The radio spectra  from these four epochs are displayed in Figure~\ref{fig:shrtlcSpind}, where we also display the corresponding light curves for reference.\footnote{For these light curves, we require $S/N>3$ for the flux densities in each time bin to reduce uncertainties on $\alpha_r$.  Since the on source integration time should be similar in each time bin, we also remove a small number of  time bins ($<$5 across all four sources) where the $\sigma_{\rm rms}$ of the time-resolved images differ by at least a factor of two from the median $\sigma_{\rm rms}$ of all images on each date.  Such large variations in $\sigma_{\rm rms}$ might  indicate that artifacts remain after the cleaning process, or that there is an unusually large amount of flagged data during that time bin.} 
     The spectral constraints during the 2003 epoch are not very meaningful, but we include that epoch in the figure for completeness.    For the other three epochs, we do not see any obvious changes of $\alpha_r$ with time;  variations of $\alpha_r$ are consistent with statistical noise, with reduced $\chi^2_r$ of 1.5, 2.5, and 1.5 (for 52, 32, and 32 degrees of freedom) on  2013 December 2, 2015 August 1, and 2015 August 5, respectively.   We note that  \citet{rana16} reported that the radio spectrum of \src\ switched from optically thick to optically thin over $\approx$10 min periods on 2013 December 2.  While we also see some variations in $\alpha_r$ over 10-30 min timescales, they tend to be at the 1--2$\sigma$ level, such that we do not consider those variations to be highly significant relative to the measurement error.  We tentatively see marginal indications for a long-term evolution of the spectrum over the  $\approx$9 hour observation.  Imaging the first 100 min of the observation yields $\alpha_r = -0.39 \pm 0.10$, while a less steep (and potentially flat) spectral index of $\alpha_r = -0.14 \pm 0.14$ is measured from images during the last 100 min.   The steeper spectral index during the first 100 min appears to be driven by variations at 7.45 GHz that are not mimicked at 5.25 GHz.

\section{Discussion}
\label{sec:disc}
We have presented radio light curves of \src\ in quiescence from \nallobs\ VLA and VLBA observations spanning 1991--2015, and we find that factor of 2--4 variations are common on timescales ranging from minutes to decades.  Eleven observations are long enough to produce light curves on sub-hour timescales,  from which we conclude that radio flares that last from tens of minutes to hours are common.  However, there is not a single template to describe quiescent radio variability, either in flare profile, amplitude, or timescale.  The observed variety in flare properties could imply that  multiple mechanisms control the radio variability, or that a single type of process  yields a range of subtly different radiative signatures (e.g., a shock traveling through a steady jet, where the dissipation of energy is highly sensitive to the local conditions at the location of the shock).    

 Only for a single large  flare (2015 August 5) do we have sufficient time resolution to attempt to statistically characterize its properties during the rise.  Its structure function displays a  slope $\beta \approx 1$, which means that at least some flares are consistent with red noise (i.e., $P(F) \propto F^{-2}$, although we note that we do not have coverage of the entire flare, which could bias our slope measurement).    Considering that the structure function of long-term variations is flat (over $\approx$10-4000 day timescales; Figure~\ref{fig:sf}), we suggest that the quiescent jet of \src\ (sometimes) displays flares with `random walk' noise characteristics that dampen to uncorrelated variations on longer timescales.  We suspect that the damping time can be as short as tens of minutes to hours, as observed during some flares in Figure~\ref{fig:shrtlc}, but from Figure~\ref{fig:sf}  we can only constrain the damping timescale to $\lesssim$10 days.

We do not assert  that all flares have red noise  characteristics, since there is only one (large) flare for which we have sufficient time resolution to calculate a structure function, and that observation was taken very shortly after the system returned to quiescence following the 2015 outburst.  Nevertheless,  the behaviour of red noise flares that dampen to uncorrelated variations on long timescales is reminiscent of decade long radio light curves of BL Lac objects, i.e., low-luminosity active galactic nuclei (AGN)  with a jet pointed toward Earth.    \citet{hughes92} find that BL Lac objects typically show first-order structure functions with slopes $\beta \sim 1$ that become flat at  long timescales (i.e., consistent with a damped random walk).  They find a broad distribution of characteristic damping timescales from $\approx$1-10 yr for BL~Lac objects.\footnote{This phenomenology is also common for jet-dominated emission in the optical waveband, where  BL~Lac object light curves can be well-characterised by a damped random walk \citep{ruan12}, and in some cases also in the gamma-ray (where sporadic flaring over a steady flicker/red noise power spectrum is often observed; e.g., \citealt{abdo10}).}  

To first order, after correcting for beaming effects related to BL~Lac orientation, we expect BL Lac objects to be analogous to hard state \xrb s \citep[e.g.,][]{falcke04, kording06}, and, despite their higher Eddington ratios,  we believe that BL~Lac objects also make reasonable analogs for comparison to a luminous quiescent \xrb\ like \src\ (i.e, both types of systems launch compact radio jets from a black hole fed by  inefficient accretion).  However, to  compare BL~Lac timescales to \src\  requires correction for the effects of relativistic beaming from the BL~Lac jets, which is difficult given unknown Doppler  factors and redshifts for most of  the BL~Lac objects in \citet{hughes92}.   As an extreme example, we  consider a   BL Lac object with a  Doppler factor  $\delta \approx 60$, which would correspond to a very fast  jet with bulk Lorentz factor $\Gamma \approx 50$ \citep[e.g.,][]{lister09} aligned within only 1 degree to our line of sight.  In that case, the intrinsic (i.e., rest-frame) characteristic timescales from \citet{hughes92} would be $\lesssim$600 yr (BL~Lac objects tend to have low redshifts, e.g., a median $z\sim0.33$ in \citealt{shaw13}, such that cosmological corrections will be smaller than beaming corrections).  
 
  If one were to associate the above timescale with a timescale that scales linearly with black hole mass (e.g., the light travel time across an emitting region that is comparable in size to the radio photosphere of a conical jet), then $\lesssim$600 yr would scale down to $\lesssim$10 min for \src\ (comparing the 9 $M_{\rm Sun}$ mass of \src\ to a typical $3\times10^8 M_{\rm Sun}$ mass for a BL Lac object, \citealt{plotkin11}).    We suspect that most flares from \src\ decay on timescales of at least several tens of minutes to hours, longer than expectations from the above mass scaling.       Therefore, we are either probing  physical variations in \src\ that are inaccessible for individual supermassive black holes, or it is possible for some AGN phenomenology to occur on much faster timescales than expected from (simplistic) mass-scaling arguments.  We view the latter as a  plausible explanation, particularly in light of  the existence of changing look AGN, which have accretion disks that appear to operate on much quicker timescales than expected from mass-scaling arguments \citep[e.g.,][]{noda18, dexter18}.  
   
Finally,  we note that the radio photosphere of the quiescent jet from \src\ (at GHz frequencies) is empirically constrained to be located at a distance $\lesssim$3.4 AU from the black hole \citep{miller-jones08, plotkin17}.  The size of the radio emitting region would be even smaller (e.g., by a factor of $\tan \phi$ for a conical jet, where $\phi$ is the jet opening angle; \citealt{miller-jones06}),  such that one should not expect radio variations on timescales longer than $\approx$tens of minutes to be causally connected unless the jet is very slow.    Thus, long-term variations  likely reflect the jet responding to separate disturbances, like changes in the mass accretion rate through the inner flow, or like changing intrinsic properties of the jet  (e.g., how internal energy in the jet is partitioned between particles and magnetic field).  That both the radio and X-ray display similar long-term variability characteristics (i.e., flat structure functions, see \citealt{bernardini14}) might support that both wavebands release radiative power by tapping into the same energy reservoir, as suggested by, e.g.,  \citet{malzac04}.\footnote{We note that we are probing longer timescales than considered by \citealt{malzac04}, who consider X-ray variations associated with dynamical timescales at 10-100 Schwarzschild ($\approx 0.1$ s), while in our case the X-ray and radio variations  are  over timescales longer than several days.  However, the general idea that both the radio and X-ray emission regions are responding to a common physical driver seems a reasonable interpretation.}  

\subsection{Spectral Variations}
\label{sec:disc:spind}

As noted in Section~\ref{sec:obs:spind:long}, we measure radio spectral indices ranging from $-1.5 < \alpha_r < +1.6$, with a mean $\alpha_r = 0.02$ and a standard deviation $\sigma_{\rm \alpha_r} = 0.65$.    We suspect that the large range in $\alpha_r$  can  be attributed to the poorer sensitivity  of the historical VLA, and  multifrequency data that can be offset by $\pm$30 min.  For example, a factor of two variability within  30 min would adjust $\alpha_r$ by $\pm1.3$.  We are therefore cautious not to over interpret the apparently large range of measured $\alpha_r$, which might not require a physical explanation.

The above variability-related concerns, however, do not preclude the possibility of less extreme spectral variations.   The upgraded VLA provides strictly simultaneous multifrequency observations, and the 2013 December 2 epoch ($\alpha_r = -0.26 \pm 0.05$) provides evidence that the spectral index can indeed stray negative at times (as originally pointed out by \citealt{rana16}).    We stress that   $\alpha_r=-0.26 \pm 0.05$ is inconsistent with a purely optically thin spectrum,  as expected from transient ejecta \citep[e.g.,][]{fender99, corbel04}.  Possible explanations for a mildly negative spectral index, as observed on 2013 December 2, could  include the combination of optically thick and optically thin emission  (e.g., the fading optically-thin stage of a flare superposed over a steady, flat spectrum jet),  a jet that expands more slowly than a conical jet, or a decelerating jet. Also contributing could be that  lower-frequency variations are `smeared', since they are emitted over a larger volume (farther from the black hole) compared to higher-frequency emission.  Indeed, the 2013 December 2 light curve (Figure~\ref{fig:shrtlcSpind}) displays dips at 7.45 GHz  at approximately 50 and 150 min after the start of the observation that are not mimicked at 5.25 GHz, which  likely contributes to the overall negative spectral index.

To our knowledge, the 2013 December 2 observation of \src\ is the only observation of a quiescent \xrb\ to  show a (well-measured) negative spectral index. However, comparably negative spectral indices have been measured from compact jets in the \textit{hard} state (see, e.g., \citealt{espinasse18} and references therein, although we note many of those observations also suffer from low sensitivity).   Reasonable explanations for different spectral indices in the hard state include  intrinsic differences in jet properties  \citep{espinasse18} \textit{or} differences in inclination \citep{motta18} (or both).  One difference  in our work, however, is that we are seeing both  positive and  negative spectral indices from the same source.  Therefore, we cannot explain a varying radio spectral index from \src\ in quiescence as an inclination effect.

\section{Coordinating Multiwavelength Observations}
\label{sec:mwcoord}
Part of our motivation for this work is to quantify the level to which variability-related systematic uncertainties can influence the location of quiescent BHXBs  in the radio/X-ray luminosity plane ($\lr-\lx$).   Understanding these uncertainties is important for studies on disk/jet couplings (e.g.,  fitting $\lr-\lx$ correlation slopes for  individual objects), and also  for studies that appeal to $\lr-\lx$ to identify new \xrb\ candidates.  Below we compare the radio variability of \src\ to its X-ray variability (taken from the literature), and we recommend some guidelines for reducing variability-induced systematics when coordinating and interpreting  radio/X-ray observations of quiescent \xrb\ (candidates) with luminosities comparable to \src.

\subsection{X-ray Variability in Quiescence}

 \citet{bernardini14} obtained dense X-ray monitoring of \src\ with the \textit{Neil Gehrels Swift Observatory}, taking 33 observations  over 75 days.  Their study  provides the most appropriate comparison to our long-term radio light curve(s) in Figure~\ref{fig:lc}.  They find a fractional rms variability in the X-ray of $F_{\rm var, X-ray} = 57 \pm 3$\% ($\sim$0.25 dex), and they obtain a flat structure function over timescales of $\approx$5-80 days.  These X-ray results are similar to our radio results, where we find $F_{\rm var, radio} = 54 \pm 6$\%  ($\sim$0.23 dex) and a flat structure function (although our study extends to longer timescales of of $\approx$10-4000 days).  
 
While the level of long-term X-ray variability is comparable to that in the radio, short-term X-ray variability can be larger in amplitude.  For example, on hour timescales factor of 4--8 X-ray variations are typical  \citep[e.g.,][]{bradley07, bernardini14, rana16}; at the extreme end, \citet{hynes04} observed a flare that increased  the \textit{Chandra} count rate by a factor $>$20 (also see \citealt{wagner94} for a factor of 10 variation in $<$0.5 days).  For comparison, in our work we commonly observe flares that change the flux density by factors of 2--4 in the radio. 

 Whether or not one should expect correlated variations between the radio and X-ray bands on short timescales is an open question.   To our knowledge, there have not been  multiwavelength campaigns that simultaneously take radio and X-ray observations \textit{in quiescence} over multiple epochs separated by only 1--2 days, thereby making it impossible to empirically test if radio/X-ray variations are correlated on sub-week timescales.  On minute through hour timescales, only two attempts have been published so far that searched for coordinated radio/X-ray variability over observations lasting several hours (2003 July 29 and 2013 December 2), and neither attempt has shown obvious radio/X-ray correlations with light curves on 30--100 min time bins \citep{hynes09, rana16}.  However, detecting correlated variations may require finer time resolution.

\subsection{Recommendations for Coordinating Radio and X-ray Observations}
Considering the above X-ray characteristics, and that correlated (short-term) radio and X-ray variability might only be detectable with minute time resolution, we suggest the following strategies when coordinating multiwavelength campaigns on quiescent X-ray binaries:

\begin{itemize}

\item \textit{radio and X-ray observations should be scheduled as simultaneously as possible.}  If the source is bright enough to provide sufficient signal-to-noise on $<$10 min time bins in  both the radio and X-ray, then (some) individual flares should be resolvable, and one can directly  control for multiwavelength variability.  Ideally the observations would be long enough to observe the entire flare rise and decay (which can last several hours). 

\item \textit{If data are non-simultaneous, or if one does not have sufficient time resolution to resolve individual flares, then inflate the  error bars by $\approx$0.25 dex  in both the radio and the X-ray.}  However, one should still bear in mind that variations as large as factors of 2--4 in the radio and 4--8 in the X-ray are common. 

\item  \textit{Radio observations should be taken as close as possible to the  frequency required to achieve one's science goals.}  If spectral constraints from strictly simultaneous multifrequency data are lacking, then we support the standard assumption of a flat radio spectrum when extrapolating radio observations to other frequencies.  However, we find evidence that $\alpha_r$ is not always strictly zero, and we recommend propagating errors on flux densities by assuming that the radio spectrum could vary by $\sigma_{\alpha_r} \approx \pm 0.6$ (see Section~\ref{sec:obs:spind:long}).  

\end{itemize}

\subsection{A Comparison to the Transitional Millisecond Pulsar PSR J1023+0038}
\label{sec:mwcoord:tmsp}

While quiescent \xrb s tend to have higher $\lr/\lx$ ratios than other classes of Galactic compact accreting objects  \citep[e.g.,][]{migliari06, tudor17, gallo18},  the transitional millisecond pulsar (tMSP) PSR J1023+0038 (hereafter \tmsp) was recently shown to sometimes be an exception to that trend \citep{bogdanov18}.  Considering periods when \tmsp\ does not show radio pulsations as accretion-powered states, \tmsp\ exhibits aperiodic and rapid switching between two distinct X-ray flux levels, a low mode ($\lx \approx 5 \times 10^{32}~\ergs$) and a high mode ($\lx \approx 3 \times 10^{33} ~\ergs$; e.g., \citealt{patruno14, stappers14, jaodand16, bogdanov15}).  \citet{bogdanov18} discovered that  the low modes are nearly always accompanied by radio flares that both rise and decay on minute timescales.  The radio flux density appears anti-correlated with the X-ray flux, although some radio flaring is also observed when the X-ray flux remains in a high mode.  During these low X-ray mode states, the increased radio flaring tends to reach $\lr \approx 2 \times 10^{27}\,\ergs$.  

\begin{figure*}
\begin{center}
\includegraphics[scale=0.6]{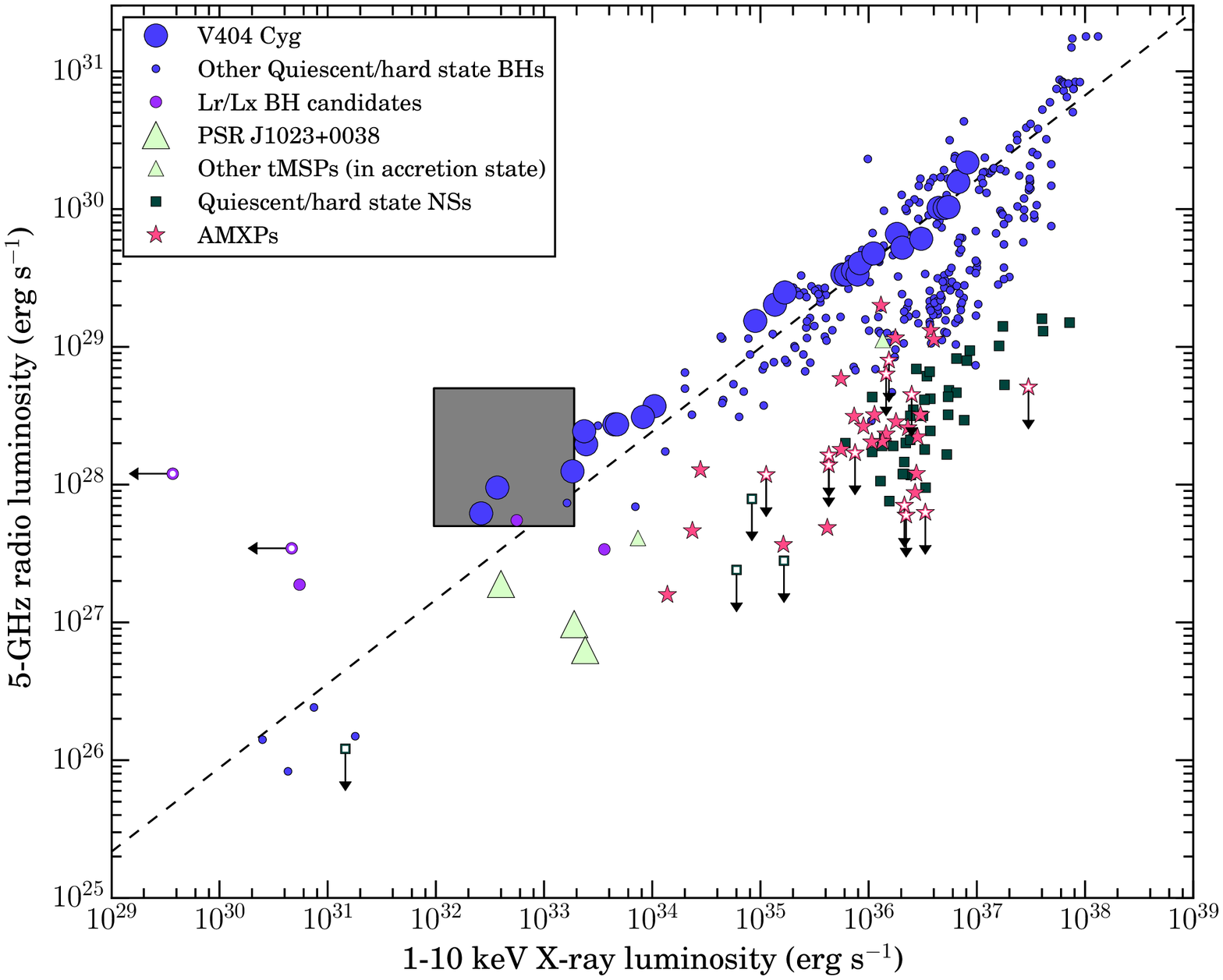}
\caption{Radio (5 GHz) vs. X-ray (1-10 keV) luminosities for quiescent and hard state \xrb s (blue circles), quiescent and hard state neutron star (NS) X-ray binaries (black squares), accreting millisecond X-ray pulsars (AMXPs, red stars), and transitional millisecond pulsars in accretion-powered states (tMSPs, green triangles).  Highlighted with large symbols are \src\ (from \citealt{hynes04, corbel08, rana16, plotkin17}) and the tMSP PSR J1023+0038 (from \citealt{deller15, bogdanov18}).  All other data points are taken   from \citet{arashlrlx}.  The gray shaded box illustrates the minimum and maximum luminosities displayed by \src\ in quiescence (radio luminosities taken from this work, X-ray luminosities taken from \citealt{bernardini14}), assuming that multiwavelength variations are uncorrelated.    At low X-ray luminosities the tMSP \tmsp\  tends to remain at least a factor of 2.5 radio fainter than \src, but it still starts entering parameter space that could  be occupied by \xrb s.} 
\end{center}
\label{fig:lrlx}
\end{figure*}

In Figure~\ref{fig:lrlx} we highlight the location of \tmsp\ in the $\lr-\lx$ plane\footnote{Data taken from \citet{arashlrlx}, see \url{https://github.com/bersavosh/XRB-LrLx_pub}.}  
 (large green triangles) compared to \src\ in quiescence and in the hard state (large blue circles).  We also shade in a region that represents the minimum and maximum luminosities displayed by \src\ within our radio dataset ($\approx 5\times10^{27} - 5\times10^{28}~\ergs$, which corresponds to flux densities ranging from 0.14-1.35 mJy if one assumes a flat radio spectrum), and the minimum and maximum X-ray luminosities displayed during the \textit{Swift} campaign from \citet{bernardini14}.  We estimate 1-10 keV X-ray luminosities assuming a photon index of 2.1 and a column density of $9\times10^{21}$ cm$^{-2}$ \citep{bernardini14}, using X-ray count rates extracted from the online \textit{Swift}-XRT product generator tool\footnote{\url{http://www.swift.ac.uk/user_objects}} \citep{evans07, evans09}.  The shaded region in Figure~\ref{fig:lrlx} assumes that radio and X-ray luminosities are not correlated in quiescence, and therefore represents a conservative range for where one might find \src\ in the $\lr-\lx$ plane if considering non-simultaneous data.  The mean radio flux density from our study ($\log f_\nu/{\rm mJy} = -0.53$) corresponds to $\log \lr/\ergs\,\approx 28$, such that \src\ likely spends most of its time toward the bottom left of the shaded region (with the upper right region representing periods of flaring activity).

 After accounting for the range of flux density variations exhibited by \src, we expect its radio luminosity to always be $\gtrsim$2.5 times larger than the radio luminosity of \tmsp, even when \tmsp\ is in a low X-ray flux (i.e., radio flaring) state.  Still, \tmsp\ ventures very close to the parameter space expected for quiescent \xrb s, and it is easy to envision that it could overlap with a \xrb\ that has similar variability characteristics as \src\ but  at a slightly lower Eddington ratio.  Although \tmsp\ only ventures toward the \xrb\ parameter space on short timescales $\lesssim$500 s (i.e., over longer time-averaged observations, \tmsp\ will generally appear radio fainter), other tMSPs could venture toward the BHXB parameter space for longer periods of time (e.g., the tMSP IGR J18245$-$2452 has displayed at least one low mode that lasted for nearly 10 hours, e.g., \citealt{linares14}).  We therefore reiterate the conclusion from \citet{bogdanov18} that radio and X-ray luminosities alone are not always sufficient to confidently assert that a quiescent  X-ray binary contains a  black hole instead of a neutron star.\footnote{Although in some cases, e.g., a system being very radio bright or radio faint, one could reasonably disfavor or favor a neutron star from radio and X-ray luminosities.} 
   Other multiwavelength data should naturally also be consulted to, e.g., search for an orbital period, and to search for emission lines or other properties to give clues on the nature of the companion star and accreting compact object \citep[e.g.,][]{bahramian17, shishkovsky18, tudor18}.      
       
 Lacking detailed multiwavelength data, we assert that radio light curves could have some diagnostic power to determine if one is observing a black hole or a neutron star: 
  
\begin{itemize}
\item \textit{frequent and relatively short radio flares that both rise and decay on rapid (minute) timescales may favor a neutron star.}   \citet{bogdanov18} find radio flares that last $\lesssim$500 s, and they rise and decay rapidly (see their Figures 1, 4, and 5).  These flares can occur quite frequently at times (e.g., \citealt{bogdanov18} observe as many as 3-4 radio flares per hour during some portions of their observation).  For \src, we observe less frequent flaring, and we only observe rapid flare rises.  The decays tend to proceed on longer  timescales of tens of minutes to hours;  

\item \textit{an order of magnitude change in the quiescent radio flux density over minute-to-hour timescales might argue for a neutron star.}  \citet{deller15} show light curves of \tmsp\ from 13 different epochs (see their Figure 5), and they find a case where the radio flux density steadily increases by an order of magnitude over $\approx$30 min.  While \src\ has also been observed to show flares with comparably slow rise times, none of those $>$30 min flares have increased in flux density by an order of magnitude. 

\end{itemize}

We stress that the above assertions  are predicated on observing a system that is close enough to produce radio light curves on minute timescales (e.g., \src\ is at $2.39\pm0.14$ kpc, \citealt{miller-jones09}, and \tmsp\ is at $1.368^{+0.042}_{-0.039}$ kpc, \citealt{deller12}).  Also, in some cases radio light curves will not hold any diagnostic power, since, particularly with shorter observations,  \tmsp\ and \src\ can both show time periods of low activity.  Furthermore, the  variability observed so far from \tmsp\ and from \src\ unlikely represents the full range of variability that can be displayed by either class of systems (e.g., as evidenced by the nearly 10 hour X-ray low mode displayed by the tMSP  IGR J18245$-$2452). 

If one does not have access to radio light curves with minute time resolution, then taking a  radio flux density integrated over long observations could provide a less ambiguous interpretation: over long  time intervals, non-flaring  time periods will dilute the radio flux density, thereby moving the tMSP farther away from the quiescent \xrb\ space in $\lr-\lx$.  Finally, in some cases the X-ray spectrum can provide additional diagnostic power:  \tmsp\ has an X-ray photon index of $\Gamma \approx 1.7$ in both low and high X-ray modes \citep{bogdanov18}, while quiescent \xrb s are well established to have a  softer $\Gamma \approx 2.1$ in quiescence \citep{plotkin13, reynolds14}. The radio spectral index, however, is a poor discriminant, given that both \src\ and \tmsp\ exhibit a range of mildly positive and negative spectral indices  (\citealt{bogdanov18} report a range extending from $-0.5 \lesssim \alpha_r \lesssim 0.4)$, and that $\alpha_r$ is usually not well-measured in low-luminosity sources.  

\section{Summary}
\label{sec:conc}

We present archival radio observations of \src\ spanning 24 years (1991--2015), providing the most stringent long-term constraints to date on the radio variability of a synchrotron jet from a quiescent  \xrb.  We find flux densities that follow a lognormal distribution, with mean and standard deviation (in $\log f_\nu/{\rm mJy}$) of  $-0.53 \pm 0.19$ and $-0.53 \pm 0.30$ at 4.9 and 8.4 GHz, respectively.  Factor of $>$2-4 variations  are common on every observable timescale from minutes to decades.   As expected, the radio spectrum of \src\ is flat on average ($\left<\alpha_r\right> = 0.02 \pm 0.65$, where the error represents the standard deviation), but we also find that $\alpha_r$ becomes significantly negative on at least one epoch.

Over  two decades of observations, \src\ displays a flat structure function, such that the long-term flux density variations  (days -- years) are consistent with either flicker noise ($P(F)\propto F^{-1}$) or white noise.     On epochs when we have sufficient quality data, we  observe individual flares   that appear to decay within minutes to hours.  These results are consistent with an interpretation of  shot noise probed over a timescale that is longer than the characteristic damping time of each disturbance.   We suspect that typical flare decay timescales are on the order of tens of minutes  to hours, but we are formally  only capable of constraining that timescale to $<$10 days (with tens of minutes a reasonable lower bound).   These  properties may be expected from shock instabilities traveling through a steady, compact jet.  Given similar characteristics observed in the X-ray band \citep{bernardini14}, the radio variability appears consistent with scenarios where the X-ray and radio radiative processes are powered by a common energy source \citep[e.g.,][]{malzac04}.   

Finally, we provide recommendations for combatting variability-induced systematics when attempting to place accreting compact objects onto the radio/X-ray luminosity plane, as is commonly done in surveys  for quiescent \xrb s.  We recommend that radio and X-ray observations be taken as simultaneously as possible to allow direct detections of flares during each observation.  If the data are not of sufficient quality to detect individual flares, or the data are not simultaneous, then we recommend inflating error bars on radio and X-ray luminosities by 0.25 dex.  If one must extrapolate radio observations to different frequencies, then ideally one would be able to measure a radio spectrum from  strictly simultaneous multi-frequency observations.  Otherwise, a flat radio spectrum is a reasonable approximation, except that error bars should also be adjusted according to expectations that the radio spectrum can vary ($\sigma_{\rm \alpha_r} = \pm 0.6$ is a reasonable uncertainty).  Finally, we repeat warnings by \citet{bogdanov18} that some accreting neutron stars (i.e., tMSPs) can obtain radio and X-ray luminosities comparable to those achieved by quiescent \xrb s, such that other types of multiwavelength data (e.g., optical, ultraviolet, and X-ray spectra and timing) should be considered when attempting to identify the nature of accreting compact objects when mass functions are not available.

\acknowledgements
We are grateful to Robert Hjellming, who obtained many of the observations presented in this paper and who lay the foundations for understanding stellar sources at radio wavelengths.  We are also grateful to Michael Rupen for obtaining a large portion of the VLA observations used in this paper.  We thank the anonymous referee for helpful comments. The National Radio Astronomy Observatory is a facility of the National Science Foundation (NSF) operated under cooperative agreement by Associated Universities, Inc.  This work made use of data supplied by the UK \textit{Swift} Science Data Centre at the University of Leicester.   We acknowledge support from  NSF grant AST-1308124.  R.M.P. acknowledges support from Curtin University through the Peter Curran Memorial Fellowship. J.C.A.M.J. is supported by an Australian Research Council Future Fellowship (FT140101082).  L.C. acknowledges support from NSF AST-1412549.  J.S. acknowledges support from the Packard Foundation.   

 \facilities{VLA, VLBA}

\software{{\tt AIPS} \citep[v31DEC2014, v31DEC2015][]{greisen03}, {\tt Astropy} \citep{astropy-collaboration13}, {\tt CASA} \citep[v5.1.1][]{mcmullin07}}



\begin{thebibliography}{}
\expandafter\ifx\csname natexlab\endcsname\relax\def\natexlab#1{#1}\fi
\providecommand{\url}[1]{\href{#1}{#1}}

\bibitem[{{Abdo} {et~al.}(2010){Abdo}, {Ackermann}, {Ajello}, {Antolini},
  {Baldini}, {Ballet}, {Barbiellini}, {Bastieri}, {Bechtol}, {Bellazzini},
  {Berenji}, {Blandford}, {Bloom}, {Bonamente}, {Borgland}, {Bouvier},
  {Bregeon}, {Brez}, {Brigida}, {Bruel}, {Buehler}, {Burnett}, {Buson},
  {Caliandro}, {Cameron}, {Caraveo}, {Carrigan}, {Casandjian}, {Cavazzuti},
  {Cecchi}, {{\c C}elik}, {Chekhtman}, {Cheung}, {Chiang}, {Ciprini}, {Claus},
  {Cohen-Tanugi}, {Cominsky}, {Conrad}, {Costamante}, {Cutini}, {Dermer}, {de
  Angelis}, {de Palma}, {Silva}, {Drell}, {Dubois}, {Dumora}, {Farnier},
  {Favuzzi}, {Fegan}, {Focke}, {Fortin}, {Frailis}, {Fukazawa}, {Funk},
  {Fusco}, {Gargano}, {Gasparrini}, {Gehrels}, {Germani}, {Giebels},
  {Giglietto}, {Giommi}, {Giordano}, {Glanzman}, {Godfrey}, {Grenier},
  {Grondin}, {Grove}, {Guiriec}, {Hadasch}, {Hayashida}, {Hays}, {Healey},
  {Horan}, {Hughes}, {Itoh}, {J{\'o}hannesson}, {Johnson}, {Johnson}, {Kamae},
  {Katagiri}, {Kataoka}, {Kawai}, {Kn{\"o}dlseder}, {Kuss}, {Lande}, {Larsson},
  {Latronico}, {Lemoine-Goumard}, {Longo}, {Loparco}, {Lott}, {Lovellette},
  {Lubrano}, {Madejski}, {Makeev}, {Massaro}, {Mazziotta}, {McEnery},
  {Michelson}, {Mitthumsiri}, {Mizuno}, {Moiseev}, {Monte}, {Monzani},
  {Morselli}, {Moskalenko}, {Mueller}, {Murgia}, {Nolan}, {Norris}, {Nuss},
  {Ohno}, {Ohsugi}, {Omodei}, {Orlando}, {Ormes}, {Ozaki}, {Panetta}, {Parent},
  {Pelassa}, {Pepe}, {Pesce-Rollins}, {Piron}, {Porter}, {Rain{\`o}}, {Rando},
  {Razzano}, {Reimer}, {Reimer}, {Ritz}, {Rodriguez}, {Romani}, {Roth}, {Ryde},
  {Sadrozinski}, {Sander}, {Scargle}, {Sgr{\`o}}, {Shaw}, {Smith}, {Spandre},
  {Spinelli}, {Starck}, {Strickman}, {Suson}, {Takahashi}, {Takahashi},
  {Tanaka}, {Thayer}, {Thayer}, {Thompson}, {Tibaldo}, {Torres}, {Tosti},
  {Tramacere}, {Uchiyama}, {Usher}, {Vasileiou}, {Vilchez}, {Vitale}, {Waite},
  {Wallace}, {Wang}, {Winer}, {Wood}, {Yang}, {Ylinen}, \& {Ziegler}}]{abdo10}
{Abdo}, A.~A., {Ackermann}, M., {Ajello}, M., {et~al.} 2010, \apj, 722, 520

\bibitem[{{Agol} \& {Kamionkowski}(2002)}]{agol02}
{Agol}, E., \& {Kamionkowski}, M. 2002, \mnras, 334, 553

\bibitem[{{Astropy Collaboration} {et~al.}(2013){Astropy Collaboration},
  {Robitaille}, {Tollerud}, {Greenfield}, {Droettboom}, {Bray}, {Aldcroft},
  {Davis}, {Ginsburg}, {Price-Whelan}, {Kerzendorf}, {Conley}, {Crighton},
  {Barbary}, {Muna}, {Ferguson}, {Grollier}, {Parikh}, {Nair}, {Unther},
  {Deil}, {Woillez}, {Conseil}, {Kramer}, {Turner}, {Singer}, {Fox}, {Weaver},
  {Zabalza}, {Edwards}, {Azalee Bostroem}, {Burke}, {Casey}, {Crawford},
  {Dencheva}, {Ely}, {Jenness}, {Labrie}, {Lim}, {Pierfederici}, {Pontzen},
  {Ptak}, {Refsdal}, {Servillat}, \& {Streicher}}]{astropy-collaboration13}
{Astropy Collaboration}, {Robitaille}, T.~P., {Tollerud}, E.~J., {et~al.} 2013,
  \aap, 558, A33

\bibitem[{{Bahramian} {et~al.}(2017){Bahramian}, {Heinke}, {Tudor},
  {Miller-Jones}, {Bogdanov}, {Maccarone}, {Knigge}, {Sivakoff}, {Chomiuk},
  {Strader}, {Garcia}, \& {Kallman}}]{bahramian17}
{Bahramian}, A., {Heinke}, C.~O., {Tudor}, V., {et~al.} 2017, \mnras, 467, 2199

\bibitem[{Bahramian {et~al.}(2018)Bahramian, Miller-Jones, Strader, Tetarenko,
  Plotkin, Rushton, Tudor, Motta, \& Shishkovsky}]{arashlrlx}
Bahramian, A., Miller-Jones, J., Strader, J., {et~al.} 2018, Zenodo,
  doi:10.5281/zenodo.1252036.
\newblock \url{{https://doi.org/10.5281/zenodo.1252036}}

\bibitem[{{Barthelmy} {et~al.}(2015){Barthelmy}, {D'Ai}, {D'Avanzo}, {Krimm},
  {Lien}, {Marshall}, {Maselli}, \& {Siegel}}]{barthelmy15}
{Barthelmy}, S.~D., {D'Ai}, A., {D'Avanzo}, P., {et~al.} 2015, GRB Coordinates
  Network, Circular Service, No.~17929, \#1 (2015), 17929

\bibitem[{{Bernardini} \& {Cackett}(2014)}]{bernardini14}
{Bernardini}, F., \& {Cackett}, E.~M. 2014, \mnras, 439, 2771

\bibitem[{{Blandford} \& {K{\"o}nigl}(1979)}]{blandford79}
{Blandford}, R.~D., \& {K{\"o}nigl}, A. 1979, \apj, 232, 34

\bibitem[{{Bogdanov} {et~al.}(2015){Bogdanov}, {Archibald}, {Bassa}, {Deller},
  {Halpern}, {Heald}, {Hessels}, {Janssen}, {Lyne}, {Mold{\'o}n}, {Paragi},
  {Patruno}, {Perera}, {Stappers}, {Tendulkar}, {D'Angelo}, \&
  {Wijnands}}]{bogdanov15}
{Bogdanov}, S., {Archibald}, A.~M., {Bassa}, C., {et~al.} 2015, \apj, 806, 148

\bibitem[{{Bogdanov} {et~al.}(2018){Bogdanov}, {Deller}, {Miller-Jones},
  {Archibald}, {Hessels}, {Jaodand}, {Patruno}, {Bassa}, \&
  {D'Angelo}}]{bogdanov18}
{Bogdanov}, S., {Deller}, A.~T., {Miller-Jones}, J.~C.~A., {et~al.} 2018, \apj,
  856, 54

\bibitem[{{Bradley} {et~al.}(2007){Bradley}, {Hynes}, {Kong}, {Haswell},
  {Casares}, \& {Gallo}}]{bradley07}
{Bradley}, C.~K., {Hynes}, R.~I., {Kong}, A.~K.~H., {et~al.} 2007, \apj, 667,
  427

\bibitem[{{Casares}(2018)}]{casares18}
{Casares}, J. 2018, \mnras, 473, 5195

\bibitem[{{Casares} {et~al.}(1992){Casares}, {Charles}, \&
  {Naylor}}]{casares92}
{Casares}, J., {Charles}, P.~A., \& {Naylor}, T. 1992, \nat, 355, 614

\bibitem[{{Chomiuk} {et~al.}(2013){Chomiuk}, {Strader}, {Maccarone},
  {Miller-Jones}, {Heinke}, {Noyola}, {Seth}, \& {Ransom}}]{chomiuk13}
{Chomiuk}, L., {Strader}, J., {Maccarone}, T.~J., {et~al.} 2013, \apj, 777, 69

\bibitem[{{Corbel} {et~al.}(2013){Corbel}, {Coriat}, {Brocksopp}, {Tzioumis},
  {Fender}, {Tomsick}, {Buxton}, \& {Bailyn}}]{corbel13}
{Corbel}, S., {Coriat}, M., {Brocksopp}, C., {et~al.} 2013, \mnras, 428, 2500

\bibitem[{{Corbel} {et~al.}(2004){Corbel}, {Fender}, {Tomsick}, {Tzioumis}, \&
  {Tingay}}]{corbel04}
{Corbel}, S., {Fender}, R.~P., {Tomsick}, J.~A., {Tzioumis}, A.~K., \&
  {Tingay}, S. 2004, \apj, 617, 1272

\bibitem[Corbel et al.(2006)]{corbel06} Corbel, S., Tomsick, J.~A., \& Kaaret, P.\ 2006, \apj, 636, 971 

\bibitem[{{Corbel} {et~al.}(2008){Corbel}, {Koerding}, \& {Kaaret}}]{corbel08}
{Corbel}, S., {Koerding}, E., \& {Kaaret}, P. 2008, \mnras, 389, 1697

\bibitem[{{Deller} {et~al.}(2012){Deller}, {Archibald}, {Brisken},
  {Chatterjee}, {Janssen}, {Kaspi}, {Lorimer}, {Lyne}, {McLaughlin}, {Ransom},
  {Stairs}, \& {Stappers}}]{deller12}
{Deller}, A.~T., {Archibald}, A.~M., {Brisken}, W.~F., {et~al.} 2012, \apjl,
  756, L25

\bibitem[{{Deller} {et~al.}(2015){Deller}, {Moldon}, {Miller-Jones}, {Patruno},
  {Hessels}, {Archibald}, {Paragi}, {Heald}, \& {Vilchez}}]{deller15}
{Deller}, A.~T., {Moldon}, J., {Miller-Jones}, J.~C.~A., {et~al.} 2015, \apj,
  809, 13

\bibitem[{{Dexter} \& {Begelman}(2018)}]{dexter18}
{Dexter}, J., \& {Begelman}, M.~C. 2018, ArXiv e-prints, arXiv:1807.03314

\bibitem[{{Din{\c c}er} {et~al.}(2018){Din{\c c}er}, {Bailyn}, {Miller-Jones},
  {Buxton}, \& {MacDonald}}]{dincer18}
{Din{\c c}er}, T., {Bailyn}, C.~D., {Miller-Jones}, J.~C.~A., {Buxton}, M., \&
  {MacDonald}, R.~K.~D. 2018, \apj, 852, 4

\bibitem[{{Douna} {et~al.}(2018){Douna}, {Pellizza}, {Laurent}, \&
  {Mirabel}}]{douna18}
{Douna}, V.~M., {Pellizza}, L.~J., {Laurent}, P., \& {Mirabel}, I.~F. 2018,
  \mnras, 474, 3488

\bibitem[{{Dzib} {et~al.}(2015){Dzib}, {Massi}, \& {Jaron}}]{dzib15}
{Dzib}, S.~A., {Massi}, M., \& {Jaron}, F. 2015, \aap, 580, L6

\bibitem[{{Evans} {et~al.}(2007){Evans}, {Beardmore}, {Page}, {Tyler},
  {Osborne}, {Goad}, {O'Brien}, {Vetere}, {Racusin}, {Morris}, {Burrows},
  {Capalbi}, {Perri}, {Gehrels}, \& {Romano}}]{evans07}
{Evans}, P.~A., {Beardmore}, A.~P., {Page}, K.~L., {et~al.} 2007, \aap, 469,
  379

\bibitem[{{Evans} {et~al.}(2009){Evans}, {Beardmore}, {Page}, {Osborne},
  {O'Brien}, {Willingale}, {Starling}, {Burrows}, {Godet}, {Vetere}, {Racusin},
  {Goad}, {Wiersema}, {Angelini}, {Capalbi}, {Chincarini}, {Gehrels}, {Kennea},
  {Margutti}, {Morris}, {Mountford}, {Pagani}, {Perri}, {Romano}, \&
  {Tanvir}}]{evans09}
---. 2009, \mnras, 397, 1177

\bibitem[Espinasse \& Fender(2018)]{espinasse18} Espinasse, M., \& Fender, R.\ 2018, \mnras, 473, 4122 


\bibitem[{{Falcke} {et~al.}(2004){Falcke}, {K{\"o}rding}, \&
  {Markoff}}]{falcke04}
{Falcke}, H., {K{\"o}rding}, E., \& {Markoff}, S. 2004, \aap, 414, 895

\bibitem[Feigelson \& Nelson(1985)]{feigelson85} Feigelson, E.~D., \& Nelson, P.~I.\ 1985, \apj, 293, 192 


\bibitem[{{Fender} \& {Mu{\~n}oz-Darias}(2016)}]{fender16}
{Fender}, R., \& {Mu{\~n}oz-Darias}, T. 2016, in Lecture Notes in Physics,
  Berlin Springer Verlag, Vol. 905, Lecture Notes in Physics, Berlin Springer
  Verlag, ed. F.~{Haardt}, V.~{Gorini}, U.~{Moschella}, A.~{Treves}, \&
  M.~{Colpi}, 65

\bibitem[{{Fender} {et~al.}(1999){Fender}, {Corbel}, {Tzioumis}, {McIntyre},
  {Campbell-Wilson}, {Nowak}, {Sood}, {Hunstead}, {Harmon}, {Durouchoux}, \&
  {Heindl}}]{fender99}
{Fender}, R., {Corbel}, S., {Tzioumis}, T., {et~al.} 1999, \apjl, 519, L165

\bibitem[{{Fender}(2001)}]{fender01}
{Fender}, R.~P. 2001, \mnras, 322, 31

\bibitem[{{Fender} {et~al.}(2003){Fender}, {Gallo}, \& {Jonker}}]{fender03}
{Fender}, R.~P., {Gallo}, E., \& {Jonker}, P.~G. 2003, \mnras, 343, L99

\bibitem[{{Fender} {et~al.}(2013){Fender}, {Maccarone}, \&
  {Heywood}}]{fender13}
{Fender}, R.~P., {Maccarone}, T.~J., \& {Heywood}, I. 2013, \mnras, 430, 1538

\bibitem[{{Gaia Collaboration} {et~al.}(2018){Gaia Collaboration}, {Brown},
  {Vallenari}, {Prusti}, {de Bruijne}, {Babusiaux}, {Bailer-Jones}, {Biermann},
  {Evans}, {Eyer}, \& et~al.}]{gaia-collaboration18}
{Gaia Collaboration}, {Brown}, A.~G.~A., {Vallenari}, A., {et~al.} 2018, \aap,
  616, A1

\bibitem[{{Gallo} {et~al.}(2018){Gallo}, {Degenaar}, \& {van den
  Eijnden}}]{gallo18}
{Gallo}, E., {Degenaar}, N., \& {van den Eijnden}, J. 2018, \mnras, 478, L132

\bibitem[{{Gallo} {et~al.}(2005){Gallo}, {Fender}, \& {Hynes}}]{gallo05}
{Gallo}, E., {Fender}, R.~P., \& {Hynes}, R.~I. 2005, \mnras, 356, 1017

\bibitem[{{Gallo} {et~al.}(2006){Gallo}, {Fender}, {Miller-Jones}, {Merloni},
  {Jonker}, {Heinz}, {Maccarone}, \& {van der Klis}}]{gallo06}
{Gallo}, E., {Fender}, R.~P., {Miller-Jones}, J.~C.~A., {et~al.} 2006, \mnras,
  370, 1351

\bibitem[{{Gallo} {et~al.}(2014){Gallo}, {Miller-Jones}, {Russell}, {Jonker},
  {Homan}, {Plotkin}, {Markoff}, {Miller}, {Corbel}, \& {Fender}}]{gallo14}
{Gallo}, E., {Miller-Jones}, J.~C.~A., {Russell}, D.~M., {et~al.} 2014, \mnras,
  445, 290

\bibitem[{{Giesers} {et~al.}(2018){Giesers}, {Dreizler}, {Husser}, {Kamann},
  {Anglada Escud{\'e}}, {Brinchmann}, {Carollo}, {Roth}, {Weilbacher}, \&
  {Wisotzki}}]{giesers18}
{Giesers}, B., {Dreizler}, S., {Husser}, T.-O., {et~al.} 2018, \mnras, 475, L15

\bibitem[{{Greisen}(2003)}]{greisen03}
{Greisen}, E.~W. 2003, in Astrophysics and Space Science Library, Vol. 285,
  Information Handling in Astronomy - Historical Vistas, ed. A.~{Heck}, 109

\bibitem[{{Han} \& {Hjellming}(1992)}]{han92}
{Han}, X., \& {Hjellming}, R.~M. 1992, \apj, 400, 304

\bibitem[{{Hughes} {et~al.}(1992){Hughes}, {Aller}, \& {Aller}}]{hughes92}
{Hughes}, P.~A., {Aller}, H.~D., \& {Aller}, M.~F. 1992, \apj, 396, 469

\bibitem[{{Hynes} {et~al.}(2009){Hynes}, {Bradley}, {Rupen}, {Gallo}, {Fender},
  {Casares}, \& {Zurita}}]{hynes09}
{Hynes}, R.~I., {Bradley}, C.~K., {Rupen}, M., {et~al.} 2009, \mnras, 399, 2239

\bibitem[{{Hynes} {et~al.}(2004){Hynes}, {Charles}, {Garcia}, {Robinson},
  {Casares}, {Haswell}, {Kong}, {Rupen}, {Fender}, {Wagner}, {Gallo}, {Eves},
  {Shahbaz}, \& {Zurita}}]{hynes04}
{Hynes}, R.~I., {Charles}, P.~A., {Garcia}, M.~R., {et~al.} 2004, \apjl, 611,
  L125

\bibitem[{{Jaodand} {et~al.}(2016){Jaodand}, {Archibald}, {Hessels},
  {Bogdanov}, {D'Angelo}, {Patruno}, {Bassa}, \& {Deller}}]{jaodand16}
{Jaodand}, A., {Archibald}, A.~M., {Hessels}, J.~W.~T., {et~al.} 2016, \apj,
  830, 122

\bibitem[{{Jonker} {et~al.}(2014){Jonker}, {Torres}, {Hynes}, {Maccarone},
  {Steeghs}, {Greiss}, {Britt}, {Wu}, {Johnson}, {Nelemans}, \&
  {Heinke}}]{jonker14}
{Jonker}, P.~G., {Torres}, M.~A.~P., {Hynes}, R.~I., {et~al.} 2014, \apjs, 210,
  18

\bibitem[{{Kelly} {et~al.}(2009){Kelly}, {Bechtold}, \&
  {Siemiginowska}}]{kelly09}
{Kelly}, B.~C., {Bechtold}, J., \& {Siemiginowska}, A. 2009, \apj, 698, 895

\bibitem[{{Khargharia} {et~al.}(2010){Khargharia}, {Froning}, \&
  {Robinson}}]{khargharia10}
{Khargharia}, J., {Froning}, C.~S., \& {Robinson}, E.~L. 2010, \apj, 716, 1105

\bibitem[Kimura et al.(2017)]{kimura17} Kimura, M., Kato, T., Isogai, K., et al.\ 2017, \mnras, 471, 373 


\bibitem[Kong et al.(2002)]{kong02} Kong, A.~K.~H., McClintock, J.~E., Garcia, M.~R., Murray, S.~S., \& Barret, D.\ 2002, \apj, 570, 277 


\bibitem[{{K{\"o}rding} {et~al.}(2006){K{\"o}rding}, {Jester}, \&
  {Fender}}]{kording06}
{K{\"o}rding}, E.~G., {Jester}, S., \& {Fender}, R. 2006, \mnras, 372, 1366

\bibitem[{{Kuulkers} {et~al.}(2015){Kuulkers}, {Motta}, {Kajava}, {Homan},
  {Fender}, \& {Jonker}}]{kuulkers15}
{Kuulkers}, E., {Motta}, S., {Kajava}, J., {et~al.} 2015, The Astronomer's
  Telegram, 7647

\bibitem[{{Linares} {et~al.}(2014){Linares}, {Bahramian}, {Heinke}, {Wijnands},
  {Patruno}, {Altamirano}, {Homan}, {Bogdanov}, \& {Pooley}}]{linares14}
{Linares}, M., {Bahramian}, A., {Heinke}, C., {et~al.} 2014, \mnras, 438, 251

\bibitem[{{Lister} {et~al.}(2009){Lister}, {Cohen}, {Homan}, {Kadler},
  {Kellermann}, {Kovalev}, {Ros}, {Savolainen}, \& {Zensus}}]{lister09}
{Lister}, M.~L., {Cohen}, M.~H., {Homan}, D.~C., {et~al.} 2009, \aj, 138, 1874

\bibitem[{{Ma} {et~al.}(1998){Ma}, {Arias}, {Eubanks}, {Fey}, {Gontier},
  {Jacobs}, {Sovers}, {Archinal}, \& {Charlot}}]{ma98}
{Ma}, C., {Arias}, E.~F., {Eubanks}, T.~M., {et~al.} 1998, \aj, 116, 516

\bibitem[{{Maccarone}(2005)}]{maccarone05}
{Maccarone}, T.~J. 2005, \mnras, 360, L30

\bibitem[{{Makino}(1989)}]{makino89}
{Makino}, F. 1989, \iaucirc, 4782

\bibitem[Malyshev et al.(2015)]{malyshev15} Malyshev, D., Savchenko, V., Ferrigno, C., Bozzo, E., \& Kuulkers, E.\ 2015, The Astronomer's Telegram, 8458,  


\bibitem[{{Malzac} {et~al.}(2004){Malzac}, {Merloni}, \& {Fabian}}]{malzac04}
{Malzac}, J., {Merloni}, A., \& {Fabian}, A.~C. 2004, \mnras, 351, 253

\bibitem[{{McMullin} {et~al.}(2007){McMullin}, {Waters}, {Schiebel}, {Young},
  \& {Golap}}]{mcmullin07}
{McMullin}, J.~P., {Waters}, B., {Schiebel}, D., {Young}, W., \& {Golap}, K.
  2007, in Astronomical Society of the Pacific Conference Series, Vol. 376,
  Astronomical Data Analysis Software and Systems XVI, ed. R.~A. {Shaw},
  F.~{Hill}, \& D.~J. {Bell}, 127

\bibitem[{{Meier}(2001)}]{meier01}
{Meier}, D.~L. 2001, \apjl, 548, L9

\bibitem[{{Migliari} \& {Fender}(2006)}]{migliari06}
{Migliari}, S., \& {Fender}, R.~P. 2006, \mnras, 366, 79

\bibitem[Miller-Jones et al.(2006)]{miller-jones06} Miller-Jones, J.~C.~A., Fender, R.~P., \& Nakar, E.\ 2006, \mnras, 367, 1432 


\bibitem[{{Miller-Jones} {et~al.}(2008){Miller-Jones}, {Gallo}, {Rupen},
  {Mioduszewski}, {Brisken}, {Fender}, {Jonker}, \&
  {Maccarone}}]{miller-jones08}
{Miller-Jones}, J.~C.~A., {Gallo}, E., {Rupen}, M.~P., {et~al.} 2008, \mnras,
  388, 1751

\bibitem[{{Miller-Jones} {et~al.}(2009){Miller-Jones}, {Jonker}, {Dhawan},
  {Brisken}, {Rupen}, {Nelemans}, \& {Gallo}}]{miller-jones09}
{Miller-Jones}, J.~C.~A., {Jonker}, P.~G., {Dhawan}, V., {et~al.} 2009, \apjl,
  706, L230

\bibitem[{{Miller-Jones} {et~al.}(2015){Miller-Jones}, {Strader}, {Heinke},
  {Maccarone}, {van den Berg}, {Knigge}, {Chomiuk}, {Noyola}, {Russell},
  {Seth}, \& {Sivakoff}}]{miller-jones15}
{Miller-Jones}, J.~C.~A., {Strader}, J., {Heinke}, C.~O., {et~al.} 2015,
  \mnras, 453, 3918
  
  \bibitem[Motta et al.(2018)]{motta18} Motta, S.~E., Casella, P., \& Fender, R.~P.\ 2018, \mnras, 478, 5159 

\bibitem[Mu{\~n}oz-Darias et al.(2017)]{munoz-darias17} Mu{\~n}oz-Darias, T., Casares, J., Mata S{\'a}nchez, D., et al.\ 2017, \mnras, 465, L124 


\bibitem[{{Negoro} {et~al.}(2015){Negoro}, {Matsumitsu}, {Mihara}, {Serino},
  {Matsuoka}, {Nakahira}, {Ueno}, {Tomida}, {Kimura}, {Ishikawa}, {Nakagawa},
  {Sugizaki}, {Shidatsu}, {Sugimoto}, {Takagi}, {Kawai}, {Yoshii}, {Tachibana},
  {Yoshida}, {Sakamoto}, {Kawakubo}, {Ohtsuki}, {Tsunemi}, {Imatani},
  {Nakajima}, {Tanaka}, {Ueda}, {Kawamuro}, {Hori}, {Tsuboi}, {Kanetou},
  {Yamauchi}, {Itoh}, {Yamaoka}, \& {Morii}}]{negoro15}
{Negoro}, H., {Matsumitsu}, T., {Mihara}, T., {et~al.} 2015, The Astronomer's
  Telegram, 7646

\bibitem[{{Noda} \& {Done}(2018)}]{noda18}
{Noda}, H., \& {Done}, C. 2018, \mnras, 480, 3898

\bibitem[{{Patruno} {et~al.}(2014){Patruno}, {Archibald}, {Hessels},
  {Bogdanov}, {Stappers}, {Bassa}, {Janssen}, {Kaspi}, {Tendulkar}, \&
  {Lyne}}]{patruno14}
{Patruno}, A., {Archibald}, A.~M., {Hessels}, J.~W.~T., {et~al.} 2014, \apjl,
  781, L3

\bibitem[{{Perley} \& {Butler}(2013)}]{perley13}
{Perley}, R.~A., \& {Butler}, B.~J. 2013, \apjs, 204, 19

\bibitem[{{Plotkin} {et~al.}(2013){Plotkin}, {Gallo}, \& {Jonker}}]{plotkin13}
{Plotkin}, R.~M., {Gallo}, E., \& {Jonker}, P.~G. 2013, \apj, 773, 59

\bibitem[{{Plotkin} {et~al.}(2011){Plotkin}, {Markoff}, {Trager}, \&
  {Anderson}}]{plotkin11}
{Plotkin}, R.~M., {Markoff}, S., {Trager}, S.~C., \& {Anderson}, S.~F. 2011,
  \mnras, 413, 805

\bibitem[{{Plotkin} {et~al.}(2017){Plotkin}, {Miller-Jones}, {Gallo}, {Jonker},
  {Homan}, {Tomsick}, {Kaaret}, {Russell}, {Heinz}, {Hodges-Kluck}, {Markoff},
  {Sivakoff}, {Altamirano}, \& {Neilsen}}]{plotkin17}
{Plotkin}, R.~M., {Miller-Jones}, J.~C.~A., {Gallo}, E., {et~al.} 2017, \apj,
  834, 104

\bibitem[{{Rana} {et~al.}(2016){Rana}, {Loh}, {Corbel}, {Tomsick},
  {Chakrabarty}, {Walton}, {Barret}, {Boggs}, {Christensen}, {Craig}, {Fuerst},
  {Gandhi}, {Grefenstette}, {Hailey}, {Harrison}, {Madsen}, {Rahoui}, {Stern},
  {Tendulkar}, \& {Zhang}}]{rana16}
{Rana}, V., {Loh}, A., {Corbel}, S., {et~al.} 2016, \apj, 821, 103

\bibitem[{{Remillard} \& {McClintock}(2006)}]{remillard06}
{Remillard}, R.~A., \& {McClintock}, J.~E. 2006, \araa, 44, 49

\bibitem[{{Reynolds} {et~al.}(2014){Reynolds}, {Reis}, {Miller}, {Cackett}, \&
  {Degenaar}}]{reynolds14}
{Reynolds}, M.~T., {Reis}, R.~C., {Miller}, J.~M., {Cackett}, E.~M., \&
  {Degenaar}, N. 2014, \mnras, 441, 3656

\bibitem[{{Rib{\'o}} {et~al.}(2017){Rib{\'o}}, {Munar-Adrover}, {Paredes},
  {Marcote}, {Iwasawa}, {Mold{\'o}n}, {Casares}, {Migliari}, \&
  {Paredes-Fortuny}}]{ribo17}
{Rib{\'o}}, M., {Munar-Adrover}, P., {Paredes}, J.~M., {et~al.} 2017, \apjl,
  835, L33

\bibitem[{{Romero} {et~al.}(2017){Romero}, {Boettcher}, {Markoff}, \&
  {Tavecchio}}]{romero17}
{Romero}, G.~E., {Boettcher}, M., {Markoff}, S., \& {Tavecchio}, F. 2017, \ssr,
  207, 5

\bibitem[{{Ruan} {et~al.}(2012){Ruan}, {Anderson}, {MacLeod}, {Becker},
  {Burnett}, {Davenport}, {Ivezi{\'c}}, {Kochanek}, {Plotkin}, {Sesar}, \&
  {Stuart}}]{ruan12}
{Ruan}, J.~J., {Anderson}, S.~F., {MacLeod}, C.~L., {et~al.} 2012, \apj, 760,
  51

\bibitem[{{Shaw} {et~al.}(2013){Shaw}, {Romani}, {Cotter}, {Healey},
  {Michelson}, {Readhead}, {Richards}, {Max-Moerbeck}, {King}, \&
  {Potter}}]{shaw13}
{Shaw}, M.~S., {Romani}, R.~W., {Cotter}, G., {et~al.} 2013, \apj, 764, 135

\bibitem[{{Shishkovsky} {et~al.}(2018){Shishkovsky}, {Strader}, {Chomiuk},
  {Bahramian}, {Tremou}, {Li}, {Salinas}, {Tudor}, {Miller-Jones}, {Maccarone},
  {Heinke}, \& {Sivakoff}}]{shishkovsky18}
{Shishkovsky}, L., {Strader}, J., {Chomiuk}, L., {et~al.} 2018, \apj, 855, 55

\bibitem[{{Stappers} {et~al.}(2014){Stappers}, {Archibald}, {Hessels}, {Bassa},
  {Bogdanov}, {Janssen}, {Kaspi}, {Lyne}, {Patruno}, {Tendulkar}, {Hill}, \&
  {Glanzman}}]{stappers14}
{Stappers}, B.~W., {Archibald}, A.~M., {Hessels}, J.~W.~T., {et~al.} 2014,
  \apj, 790, 39

\bibitem[{{Strader} {et~al.}(2012){Strader}, {Chomiuk}, {Maccarone},
  {Miller-Jones}, \& {Seth}}]{strader12}
{Strader}, J., {Chomiuk}, L., {Maccarone}, T.~J., {Miller-Jones}, J.~C.~A., \&
  {Seth}, A.~C. 2012, \nat, 490, 71

\bibitem[Tetarenko et al.(2019)]{tetarenko19} Tetarenko, A.~J., Sivakoff, G.~R., Miller-Jones, J.~C.~A., et al.\ 2019, \mnras, 482, 2950 

\bibitem[{{Tetarenko} {et~al.}(2016){Tetarenko}, {Bahramian}, {Arnason},
  {Miller-Jones}, {Repetto}, {Heinke}, {Maccarone}, {Chomiuk}, {Sivakoff},
  {Strader}, {Kirsten}, \& {Vlemmings}}]{tetarenko16}
{Tetarenko}, B.~E., {Bahramian}, A., {Arnason}, R.~M., {et~al.} 2016, \apj,
  825, 10

\bibitem[{{Thompson} {et~al.}(2018){Thompson}, {Kochanek}, {Stanek}, {Badenes},
  {Post}, {Jayasinghe}, {Latham}, {Bieryla}, {Esquerdo}, {Berlind}, {Calkins},
  {Tayar}, {Johnson}, {Holoien}, {Auchettl}, \& {Covey}}]{thompson18}
{Thompson}, T.~A., {Kochanek}, C.~S., {Stanek}, K.~Z., {et~al.} 2018, ArXiv
  e-prints, arXiv:1806.02751

\bibitem[Tomsick et al.(2001)]{tomsick01} Tomsick, J.~A., Corbel, S., \& Kaaret, P.\ 2001, \apj, 563, 229 


\bibitem[{{Tudor} {et~al.}(2017){Tudor}, {Miller-Jones}, {Patruno}, {D'Angelo},
  {Jonker}, {Russell}, {Russell}, {Bernardini}, {Lewis}, {Deller}, {Hessels},
  {Migliari}, {Plotkin}, {Soria}, \& {Wijnands}}]{tudor17}
{Tudor}, V., {Miller-Jones}, J.~C.~A., {Patruno}, A., {et~al.} 2017, \mnras,
  470, 324

\bibitem[{{Tudor} {et~al.}(2018){Tudor}, {Miller-Jones}, {Knigge}, {Maccarone},
  {Tauris}, {Bahramian}, {Chomiuk}, {Heinke}, {Sivakoff}, {Strader}, {Plotkin},
  {Soria}, {Albrow}, {Anderson}, {van den Berg}, {Bernardini}, {Bogdanov},
  {Britt}, {Russell}, \& {Zurek}}]{tudor18}
{Tudor}, V., {Miller-Jones}, J.~C.~A., {Knigge}, C., {et~al.} 2018, \mnras,
  476, 1889

\bibitem[{{Vaughan} {et~al.}(2003){Vaughan}, {Edelson}, {Warwick}, \&
  {Uttley}}]{vaughan03}
{Vaughan}, S., {Edelson}, R., {Warwick}, R.~S., \& {Uttley}, P. 2003, \mnras,
  345, 1271

\bibitem[{{Wagner} {et~al.}(1994){Wagner}, {Starrfield}, {Hjellming}, {Howell},
  \& {Kreidl}}]{wagner94}
{Wagner}, R.~M., {Starrfield}, S.~G., {Hjellming}, R.~M., {Howell}, S.~B., \&
  {Kreidl}, T.~J. 1994, \apjl, 429, L25

\bibitem[{{Younes}(2015)}]{younes15}
{Younes}, G. 2015, GRB Coordinates Network, Circular Service, No.~17932, \#1
  (2015), 17932

\bibitem[{{Yuan} \& {Cui}(2005)}]{yuan05}
{Yuan}, F., \& {Cui}, W. 2005, \apj, 629, 408

\end{thebibliography}
\end{document}